\documentclass[aps,prl,twocolumn,superscriptaddress,notitlepage]{revtex4-2}

\usepackage{graphicx}
\usepackage{amsmath}
\usepackage{amssymb}
\usepackage{epstopdf}
\usepackage{braket}
\usepackage{xcolor}
\usepackage{color}
\usepackage[colorlinks,linkcolor=blue,urlcolor=blue,citecolor=blue,anchorcolor=blue]{hyperref}
\usepackage{enumitem}
\newcommand{\beginsupplement}{
    \setcounter{section}{0}
    \renewcommand{\thesection}{S\arabic{section}}
    \setcounter{equation}{0}
    \renewcommand{\theequation}{S\arabic{equation}}
    \setcounter{table}{0}
    \renewcommand{\thetable}{S\arabic{table}}
    \setcounter{figure}{0}
    \renewcommand{\thefigure}{S\arabic{figure}}
    \newcounter{SIfig}
    \renewcommand{\theSIfig}{S\arabic{SIfig}}}

\begin{document}

\title{Experimental reconstruction of the few-photon nonlinear scattering matrix from a single quantum dot in a nanophotonic waveguide}

\author{Hanna Le Jeannic}
\email{hanna.lejeannic@nbi.ku.dk}
\affiliation{Center for Hybrid Quantum Networks (Hy-Q), Niels Bohr Institute, University of Copenhagen, Blegdamsvej 17, DK-2100 Copenhagen, Denmark}
\author{Tom\'{a}s Ramos}
\email{t.ramos.delrio@gmail.com}
\affiliation{Instituto de F\'{i}sica Fundamental IFF-CSIC, Calle Serrano 113b, Madrid 28006, Spain}
\affiliation{DAiTA Lab, Facultad de Estudios Interdisciplinarios, Universidad Mayor, Santiago, Chile}
\author{Signe F. Simonsen}
\affiliation{Center for Hybrid Quantum Networks (Hy-Q), Niels Bohr Institute, University of Copenhagen, Blegdamsvej 17, DK-2100 Copenhagen, Denmark}
\author{Tommaso Pregnolato}
\affiliation{Center for Hybrid Quantum Networks (Hy-Q), Niels Bohr Institute, University of Copenhagen, Blegdamsvej 17, DK-2100 Copenhagen, Denmark}
\author{Zhe Liu}
\affiliation{Center for Hybrid Quantum Networks (Hy-Q), Niels Bohr Institute, University of Copenhagen, Blegdamsvej 17, DK-2100 Copenhagen, Denmark}

\author{R{\"u}diger Schott}
\affiliation{Lehrstuhl f{\"u}r Angewandte Festk{\"o}rperphysik, Ruhr-Universit{\"a}t, Universit{\"a}tsstrasse 150, D-44780 Bochum, Germany}
\author{Andreas D. Wieck}
\affiliation{Lehrstuhl f{\"u}r Angewandte Festk{\"o}rperphysik, Ruhr-Universit{\"a}t, Universit{\"a}tsstrasse 150, D-44780 Bochum, Germany}
\author{Arne Ludwig}
\affiliation{Lehrstuhl f{\"u}r Angewandte Festk{\"o}rperphysik, Ruhr-Universit{\"a}t, Universit{\"a}tsstrasse 150, D-44780 Bochum, Germany}
\author{Nir Rotenberg}
\affiliation{Center for Hybrid Quantum Networks (Hy-Q), Niels Bohr Institute, University of Copenhagen, Blegdamsvej 17, DK-2100 Copenhagen, Denmark}
\author{Juan Jos\'{e} Garc\'{i}a-Ripoll}
\affiliation{Instituto de F\'{i}sica Fundamental IFF-CSIC, Calle Serrano 113b, Madrid 28006, Spain}
\author{Peter Lodahl}
\email{lodahl@nbi.ku.dk}
\affiliation{Center for Hybrid Quantum Networks (Hy-Q), Niels Bohr Institute, University of Copenhagen, Blegdamsvej 17, DK-2100 Copenhagen, Denmark}

\begin{abstract}
Coherent photon-emitter interfaces offer a way to mediate efficient nonlinear photon-photon interactions, much needed for quantum information processing. Here we experimentally study the case of a two-level emitter, a quantum dot, coupled to a single optical mode in a nanophotonic waveguide. We carry out few-photon transport experiments and record the statistics of the light to reconstruct the scattering matrix elements of 1- and 2-photon components. This provides direct insight to the complex nonlinear photon interaction that contains rich many-body physics.
\end{abstract}

\maketitle

An efficient and reliable photon-photon nonlinearity is a key building block for photonic quantum information processing. One approach exploits post-selection after photon interference, but is resource demanding in requiring many auxiliary photons \cite{Ladd_2010}. An alternative and potentially more appealing strategy is to exploit an efficiently interfaced quantum emitter to introduce a direct photon-photon interaction \cite{Chang_2014,Lodahl_2015}. This approach requires highly efficient and coherent light-matter interaction in order to be sensitive to single quanta of light. It has been a long lasting challenge in quantum optics, and various strategies have been pursued using, e.g., Rydberg blockade interaction in atomic ensembles \cite{Tiarks_2018,Baur_2014}, single emitters in high finesse optical cavities \cite{Reiserer_2014,Najer_2019}, as well as superconducting stripline resonators \cite{Deppe_2008,Mirhosseini_2019,Lang_2011}. Recently, significant progress has been achieved in photonic waveguides including hollow fibers \cite{Bajcsy_2009} or nanofibers coupled to atomics ensembles \cite{Prasad_2019}, or with single atoms \cite{Goban_2014} and single solid-state emitters \cite{Javadi_2015,Nielsen_2018,Hallett_2018,Foster2019,Sipahigil_2016}.
In particular, solid-state quantum dots (QDs) in nanophotonics structures constitute a mature platform where scalable coherent single-photon sources \cite{Uppu_2020,Wang_2019} have been developed, at the core of the implementation of quantum information processing \cite{Lodahl_2017}.

QDs coupled to photonic-crystal waveguides provide a sophisticated platform to study few-photon interactions and quantum nonlinear optics~\cite{Chang_2014,chang_colloquium:_2018}. These devices can exhibit near-unity light-matter coupling efficiency ($\beta \geq 0.98 $~\cite{Arcari_2014}) and nonlinear interaction sensitive at the level of single photons~\cite{Javadi_2015,Hallett_2018}. Moreover, near-transform-limited emission lines have been demonstrated with QDs \cite{Kuhlmann_2015} and recently also in photonic-crystal waveguides \cite{Pedersen_2020}. The combination of unity coupling efficiency and low dephasing enables the deterministic scattering of few photons by a QD operating as a two-level emitter (TLE). The study of such scattering problems is a blooming area in quantum nonlinear optics \cite{roy_colloquium:_2017,sanchez-burillo_nonlinear_2015}. For the theoretical description, a range of new methods have been developed using the Bethe ansatz \cite{Shen_2007multi,Shen_2010}, input-output theory \cite{Fan_2010,das_photon_2018,Ramos_2018,trivedi_few-photon_2018}, Lippmann-Schwinger formalism \cite{zheng_waveguide_2010,huang_controlling_2013,lee_few-photon_2015}, wavefunction ansatz \cite{das_wave-function_2019}, diagrammatic methods \cite{pedersen_few-photon_2017,lang_non-equilibrium_2018}, path integrals \cite{shi_multiphoton-scattering_2015}, and polaron ansatz \cite{shi_ultrastrong_2018,bera_generalized_2014}.
Deterministic photon scattering processes have important applications for the realization of efficient photon sorting and deterministic Bell-state analyzers \cite{Witthaut_2012,Ralph_2015}, and can induce intricate many-body phenomena including strong photon correlations and complex photon bound states \cite{Shen_2007multi,Shen_2010,firstenberg_attractive_2013,Mahmoodian_2019}.
Despite this extensive theoretical work, it remains an open problem to isolate and characterize photon-photon interactions in the laboratory, and to develop and implement few- or multi-photon tomographic reconstruction techniques \cite{Ramos_2017} beyond the single-photon regime.

\begin{figure}[!t]
\centering
\includegraphics[width=7.5 cm]{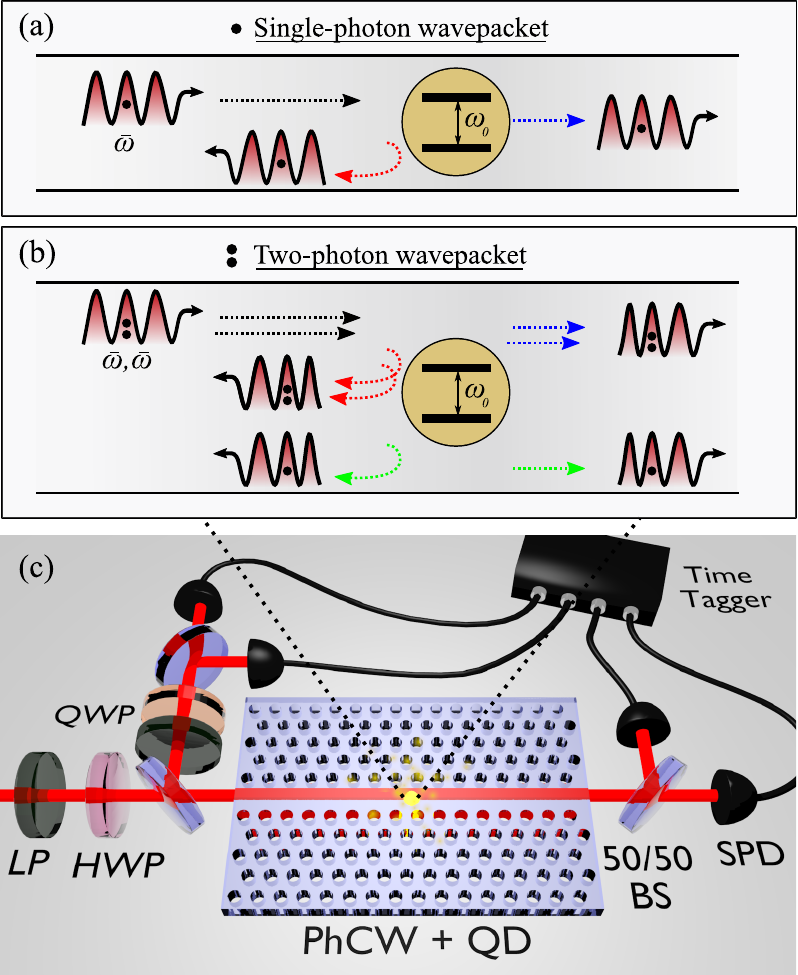}
\caption{(a,b) Illustration of single- and two-photon scattering processes for a TLE in a waveguide. In the former case, either elastic reflection or transmission may occur. In the latter case, the two photons may exchange energy via the interaction with the TLE, leading to different scattering processes. (c) Experimental setup to extract the few-photon scattering matrices of the system from intensity $I_t$ and photon correlation measurements $g_{tt}^{(2)}$, $g_{rr}^{(2)}$, and $g_{tr}^{(2)}$ between different scattering channels. The quantum dot (QD) is embedded in a photonic crystal waveguide (PhCW), and excited by a cw laser source. Beam-splitters (BS), single-photon detectors (SPD), and an electronic time tagger are used to record second-order photon correlations. Polarizing optical elements, such as two linear polarizers (LP), a half- and a quarter-wave plates (HWP and QWP, respectively) are used for extinction of the laser light and collection of the reflected light from the QD.}
\label{Figure1}
\end{figure}

This work presents the experimental reconstruction of the few-photon scattering processes of a QD in a photonic-crystal waveguide. While the coherent scattering of single photons from a TLE is simple -- the photons are either elastically reflected or transmitted [cf.~Fig.~\ref{Figure1}(a) and Ref.~\cite{Turschmann_2019}] -- the two-photon scattering processes are much more complex. In this case, different combinations of photon reflections and transmissions are possible, as shown in Fig.~\ref{Figure1}(b). Furthermore, the scattered photons become highly correlated from the interaction with the TLE \cite{Shen_2010}. We unravel these scattering processes by recording the photon statistics of the transmission and reflection outputs of a QD-waveguide system in the continuous wave (cw) regime [cf.~Fig.~\ref{Figure1}(c)]. Even in cw operation, photon-photon interactions can be extracted from photon correlation measurements, despite the low probability that incoming photons overlap in time. Our method extends a previous theoretical proposal of reconstructing multi-photon scattering properties \cite{Ramos_2017}, where the extension to photo-correlation measurements makes it insensitive to any off-chip coupling and detection inefficiencies.

Consider first for simplicity the case of a coherent TLE without dephasing that is symmetrically coupled to a waveguide (i.e.~no chiral coupling \cite{Lodahl2017_Chiral,ramos_quantum_2014}). Here, an incoming single-photon wavepacket may scatter from this system producing an output state,
\begin{align}
    |\Psi_{\rm out}^{(1)}\rangle={}&\sum_{\mu=t,r}\int d\omega f_{\bar{\omega}}(\omega)\chi^{\mu}(\omega)|1_\omega^\mu\rangle,\label{wavefunctOne}
\end{align}
that is a superposition of single-photon Fock states $\left| 1_{\omega}^\mu\right>$ of frequency $\omega$, which are either transmitted ($\mu=t$) or reflected ($\mu=r$) [cf.~Fig.~\ref{Figure1}(a)]. Both contributions are weighted by the incoming wavepacket profile $f_{\bar{\omega}}(\omega)$, centered around frequency $\bar{\omega}$, and by the single-photon transmission ($\chi^t(\omega):=t(\omega)$) and reflection ($\chi^r(\omega):=r(\omega)$) scattering coefficients, given by 
\begin{equation}\label{eq:t}
t(\omega) = 1+r(\omega) ,\quad r(\omega)=\frac{-\beta\gamma_{tot}}{\gamma_{tot}-2i\left( \omega - \omega_0 \right)}.
\end{equation}
Here, $\omega-\omega_0$ is the TLE-photon detuning, $\beta=\gamma_{wg}/\gamma_{tot}$ the waveguide coupling efficiency, and $\gamma_{tot} = \gamma_{wg}+\gamma_{l}$ the total TLE decay rate, which includes the TLE-waveguide coupling rate $\gamma_{wg}$ and the loss into unguided modes $\gamma_l$. As pointed out in Ref.~\onlinecite{Ramos_2018}, the complex coefficients $\chi_\omega^\mu$ fully characterize the single-photon scattering and satisfy $t(\omega) = 1 + r(\omega)$ even in the presence of loss or dephasing. In contrast, the intensity at transmission $I_t\geq 0$ and reflection $I_r\geq 0$ quantify the output flux and satisfy $I_t+I_r\leq 1$, where the equality only holds in the absence of decoherence channels.

Nonlinear quantum optical effects arise when an incoming two-photon wavepacket scatters off the TLE. Here, two incoming photons with frequencies $\omega_1,\omega_2$ may exchange energy via the TLE, and thereby become correlated outgoing photons with frequencies $\nu_1,\nu_2$. This process can happen either when both photons are transmitted, both reflected, or one transmitted/one reflected [cf.~Fig.~\ref{Figure1}(b)], and thus the output state $|\Psi_{\rm out}^{(2)}\rangle$ is a superposition of all these possibilities,~i.e.~\cite{Fan_2010} 
\begin{align}
    |\Psi_{\rm out}^{(2)}\rangle&={}\sum_{\mu,\mu'=t,r}\iint d\omega_1 d\omega_2 \frac{1}{\sqrt{2}} f_{\bar{\omega}}(\omega_1)f_{\bar{\omega}}(\omega_2)\label{wavefunctions}\\
    {}&\times\iint d\nu_1d\nu_2 \bigg\lbrace \chi^{\mu}(\omega_1) \chi^{\mu'}(\omega_2)\delta(\nu_2-\omega_2)\delta(\nu_1-\omega_1)\nonumber\\
    {}&+\frac{1}{2}T_{\nu_1\nu_2\omega_1\omega_2}\delta(\nu_1+\nu_2-\omega_1-\omega_2)\bigg\rbrace|1_{\nu_1}^\mu\rangle |1_{\nu_2}^{\mu'}\rangle\nonumber.
\end{align}
The first term $\sim \chi^{\mu}(\omega_1)\chi^{\mu'}(\omega_2)$ in Eq.~(\ref{wavefunctions}) corresponds to independent single-photon scattering events, and thus each photon conserves its own energy ($\nu_1=\omega_1$, $\nu_2=\omega_2$). In the last term, however, the correlated scattering coefficient $T_{\nu_1\nu_2\omega_1\omega_2}$ describes two photons acquiring a nonlinear phase shift and exchanging energy so that only the total energy is conserved ($\nu_1+\nu_2=\omega_1+\omega_2)$. For a TLE in a conventional waveguide \cite{Fan_2010}
\begin{equation}
T_{\nu_1\nu_2\omega_2\omega_1} =\frac{4}{\pi\beta\gamma_{tot}} \frac{r(\nu_1)r(\nu_2)r(\omega_1)r_(\omega_2)}{r(\frac{\omega_1+\omega_2}{2})}.\label{eq:T}
\end{equation}

This scattering matrix fully characterizes two-photon interactions including the spectral entanglement and photon-bound states induced by the TLE \cite{Shen_2007multi,Shen_2010}. The main objective in this work is to extract this information from experimental data, as described below. The analysis is generalized in the Supplemental Material (SM \cite{SM}) to include experimental imperfections including pure dephasing of the QD transition and weak Fano resonance effects.

We now describe the reconstruction protocol to experimentally characterize the nonlinear few-photon scattering processes. We follow the main idea of Ref.~\onlinecite{Ramos_2017}, which consists in illuminating the TLE with an attenuated coherent state ($|\alpha|^2\ll 1$) through one input of the waveguide. Light scatters from the TLE, creating a superposition of vacuum $|{0}\rangle$ and scattering output states for one $|\Psi_{\rm out}^{(1)}\rangle,$ two $|\Psi_{\rm out}^{(2)}\rangle$ [cf.~Eqs.~(\ref{wavefunctOne},\ref{wavefunctions})] or more photons
\begin{align}
   |\Psi_{\rm out}^{(\alpha)}\rangle = |0\rangle + \alpha |\Psi_{\rm out}^{(1)}\rangle+\frac{\alpha^2}{\sqrt{2}}|\Psi_{\rm out}^{(2)}\rangle+{\cal O}(\alpha^3).\label{weakInput}
\end{align}
The two-photon processes can be recorded in second-order correlation measurements $g^{(2)}_{\mu\mu'}$ between different output directions $\mu,\mu'$ [cf.~Fig.~\ref{Figure1}(c)]. For a monochromatic cw laser input of frequency $\omega$, we find that \cite{SM,Ramos_2020}
\begin{equation}\label{eq:g2tt}
g^{(2)}_{\mu\mu'} (\tau)=\frac{|\chi^{\mu}(\omega)\chi^{\mu'}(\omega)+\mathcal{T}(\omega,\tau)|^2}{|\chi^{\mu}(\omega)\chi^{\mu'}(\omega)|^2}+{\cal O}\left(|\alpha|^2\right).
\end{equation}
Here, $\tau$ is the time delay between the two photon detections, and ${\cal T}(\omega,\tau)$ is the Fourier transformed two-photon scattering coefficient defined as
\begin{equation}
{\cal T}(\omega,\tau)=\frac{1}{2}\int d \Delta e^{-i \Delta\tau}T_{\omega-\Delta,\omega+\Delta,\omega,\omega}.\label{FTN}
\end{equation}
The isotropy of the photon-photon interaction, i.e.~the absence of a preferred direction of emission among left and right, allows us to reconstruct experimentally the real part of this Fourier transform as \cite{SM,Ramos_2020},
\begin{align}
  {\rm Re}[{\cal T}(\omega,\tau)]=g_{tt}^{(2)}\frac{|t(\omega)|^4}{2}+g_{rr}^{(2)}\frac{|r(\omega)|^4}{2}-g_{tr}^{(2)}|t(\omega)r(\omega)|^2.\label{TwoPhotonExtraction}
\end{align}
We note that the protocol requires correlation measurements in all directions $g_{tt}^{(2)}$, $g_{rr}^{(2)}$, and $g_{tr}^{(2)}$, as well as the single-photon coefficients $t(\omega)$ and $r(\omega)$. From the real part of ${\cal T}$ we infer the imaginary part using the Kramers-Kronig (KK) relation, ${\rm Im}[\mathcal{T}(\omega,\tau)]=\frac{1}{\pi}\mathcal{P}\int d\omega'\frac{{\rm Re}[\mathcal{T}(\omega',\tau)]}{\omega-\omega'}$ \cite{SM}. Finally, an inverse Fourier transform, $T_{\omega-\Delta,\omega+\Delta,\omega,\omega}=\frac{1}{\pi}\int d\tau e^{i\Delta\tau}\mathcal{T}(\omega,\tau)$, provides the two-photon scattering matrix.

Remarkably, a measurement of the transmitted intensity $I_t(\omega)\geq 0$ suffices to extract the amplitude and phase of both complex single-photon scattering coefficients $t(\omega)$ and $r(\omega)$ \cite{Ramos_2018}. Specifically, for a weak monochromatic coherent input
\begin{equation}
I_t(\omega)=\beta-1+(2-\beta) \textrm{Re}[t(\omega)]+ {\cal O}(|\alpha|^2).\label{eq:It}
\end{equation}
From this we can infer the real part of $t(\omega)$, even in the presence of correlated dephasing noise \cite{Ramos_2018}. Using a KK relation (cf.~SM \cite{SM} for details), we then compute the imaginary part, obtaining both $t(\omega)$ and $r(\omega)=t(\omega)-1.$

We now turn to the experimental demonstration of the few-photon scattering reconstruction. We apply it to a self-assembled InGaAs QD embedded in a suspended photonic-crystal waveguide. A p-i-n diode heterostructure enables electrical contacting of the sample in order stabilize the charge environment and tune the QD. Further details of the sample wafer can be found in Ref.~\onlinecite{Kirsanske2017}. The sample is kept at T=1.6~K and a weak tunable cw laser at 938~nm (linewidth $\leq 10$~kHz, locked within a precision of 50~MHz) is launched through the waveguide via high-efficiency grating couplers \cite{SM,Zhou_2018}. Finally, the output photons are sent to superconducting nanowire single-photon detectors, with quantum efficiency $\geq 0.9$ and time-jitter below $100$~ps, to record the frequency-dependent intensity and second-order photon correlation functions.

\begin{figure}[!t]
\includegraphics[width=7 cm]{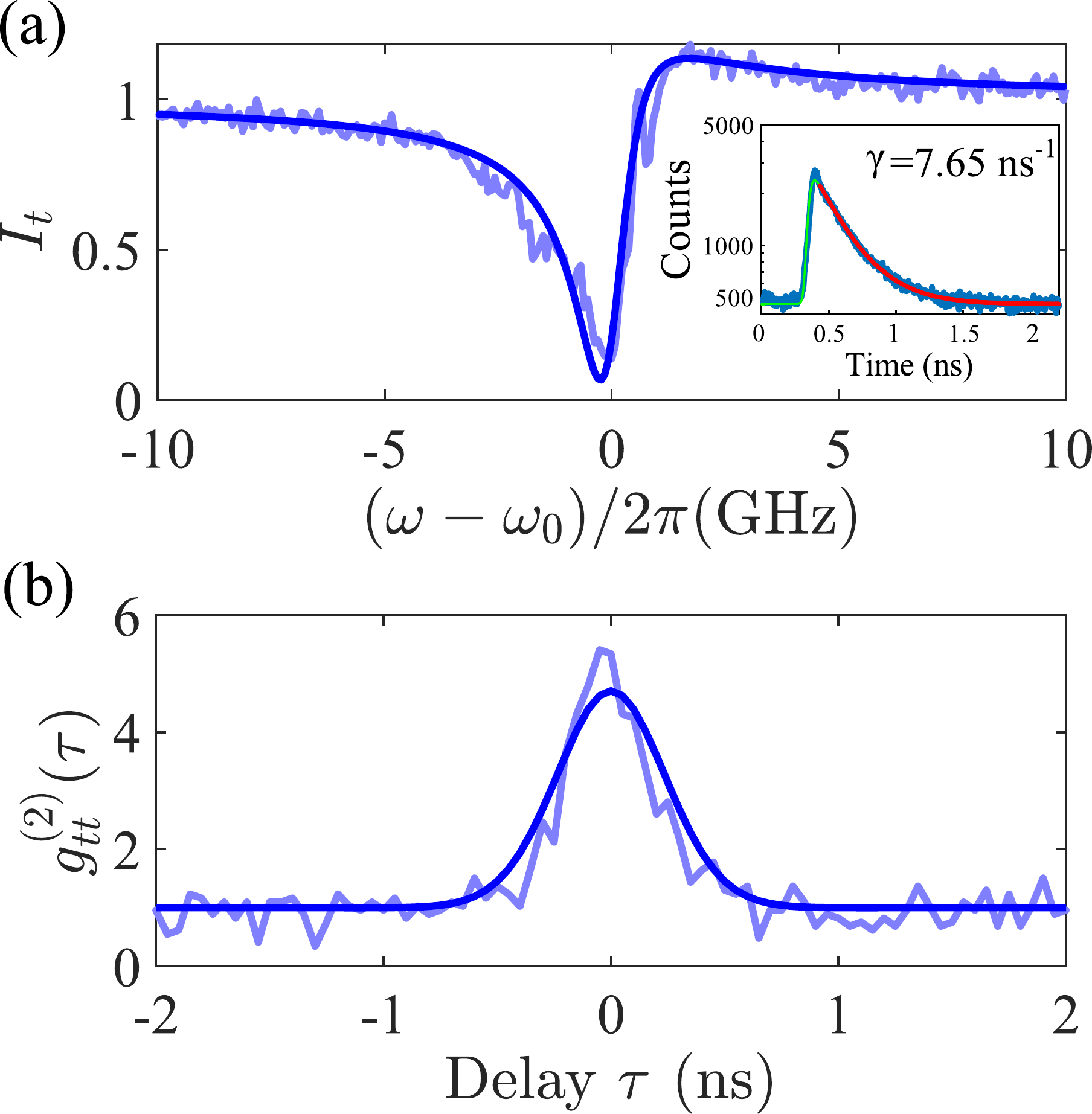}
\caption{(a) Measured (light blue) and fitted (dark blue) transmission intensity $I_t(\omega)$ as a function of the detuning of the excitation laser from the QD resonance. Inset: Time resolved dynamics of the QD (in logarithmic scale) and exponentially decaying fitting function, convolved with the instrument response of the detector, to characterize the radiative decay rate. (b) Measured (light blue) and fitted (dark blue) second-order correlation function $g_{tt}^{(2)}(\tau)$ in transmission, with time delay $\tau$, obtained on resonance ($\omega\approx\omega_0$).}
\label{Figure2}
\end{figure}

Figure~\ref{Figure2}(a) shows the transmitted intensity $I_t(\omega)$ when scanning through the QD resonance for weak excitation (on average less than $0.1$ photons per QD lifetime). The extinction of transmission on resonance exceeds $85\%$, a direct testimony of the efficiency and coherence of the photon-emitter interaction. The non-Lorentzian and asymmetric lineshape, originates from Fano resonance effects due to weak reflections at the ends of the waveguide \cite{Javadi_2015,Foster2019}. Such reflections induce a very low finesse cavity, that, depending on the overall intensity normalization (more details in SM \cite{SM}) can give $I_t$ values higher than 1. We record a transition linewidth of $\approx1.6$~GHz. For comparison the spontaneous emission decay rate is measured to be $\gamma_{tot}=\gamma_{wg}+\gamma_{l}= 7.65 \pm 0.08$ $\mathrm{ns}^{-1}$ [cf.~inset of Fig.~\ref{Figure2}(a)], corresponding to a transform-limited linewidth of $1.22$~GHz, meaning that additional broadenings due to phonons and slow spectral diffusion are less important. A full study of the statistics of QD linewidths in photonic-crystal waveguides is published elsewhere \cite{Pedersen_2020}.

Figure~\ref{Figure2}(b) shows the second-order correlation function $g^{(2)}_{tt}(\tau)$ measured in transmission, and for the same excitation conditions. It displays a pronounced bunching of $g_{tt}^{(2)}(0)\simeq 5$, which is significantly higher than in previously reported QD-waveguide experiments \cite{Javadi_2015,Hallett_2018,Foster2019} due to the substantial decoherence reduction achieved in our photon-emitter interface. The large bunching demonstrates that the incoming Poissonian photon distribution is significantly altered by the interaction with the QD, and is the experimental signature of the correlated photon-photon interaction studied in the present work.

\begin{figure}[!t]
\includegraphics[width=7.5 cm]{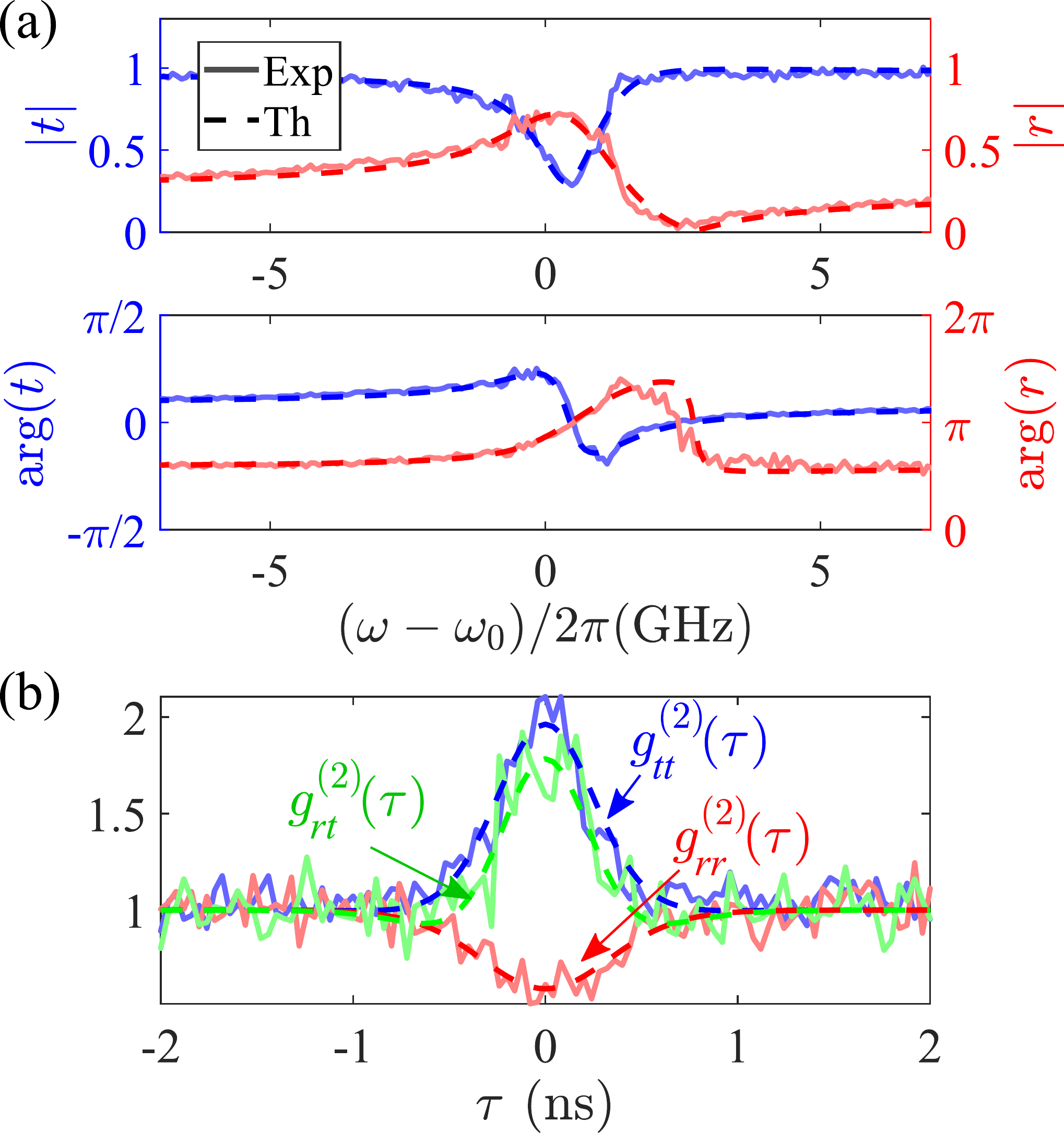}
\caption{(a)Experimentally reconstructed (solid line) modulus and phase of the complex single-photon transmission (blue) and reflection (red) coefficients and comparison to theory (dashed line). (b) Experimentally acquired second-order correlation functions for the three different configurations: $g^{(2)}_{tt}$ (blue), $g^{(2)}_{rr}$ (red), and $g^{(2)}_{rt}$ (green). The measurements are well fitted by the theoretical model including imperfections (dashed lines).}
\label{Figure3}
\end{figure}

In order to implement the two-photon reconstruction protocol, the essential governing parameters of the system must be determined first. To do so, we additionally measure the transmission intensities $I_t(\omega)$ at various excitation powers, as well as photon correlations in all directions $g_{tt}^{(2)}$, $g_{rr}^{(2)}$, and $g_{tr}^{(2)}$. 
By modelling this entire data set using a least squares fit, we arrive at a descriptive parameter set of $\beta = 0.87 \left[ 0.83,0.91 \right] $ and dephasing rate $\gamma_{d}\simeq 0 \left[ 0,0.02\right]\gamma_{tot}$, consistent with results from the literature \cite{Arcari_2014,Foster2019}. In the analysis, we also include the finite detector response time, residual spectral diffusion of the QD, background emission stemming from imperfect laser extinction or blinking of the QD state, and minor Fano resonance effects. 
We define the error ranges presented above as the $95\%$ confidence interval of each fitted parameter (more details in SM \cite{SM}).

\begin{figure}[!t]
\includegraphics[width=6.5 cm]{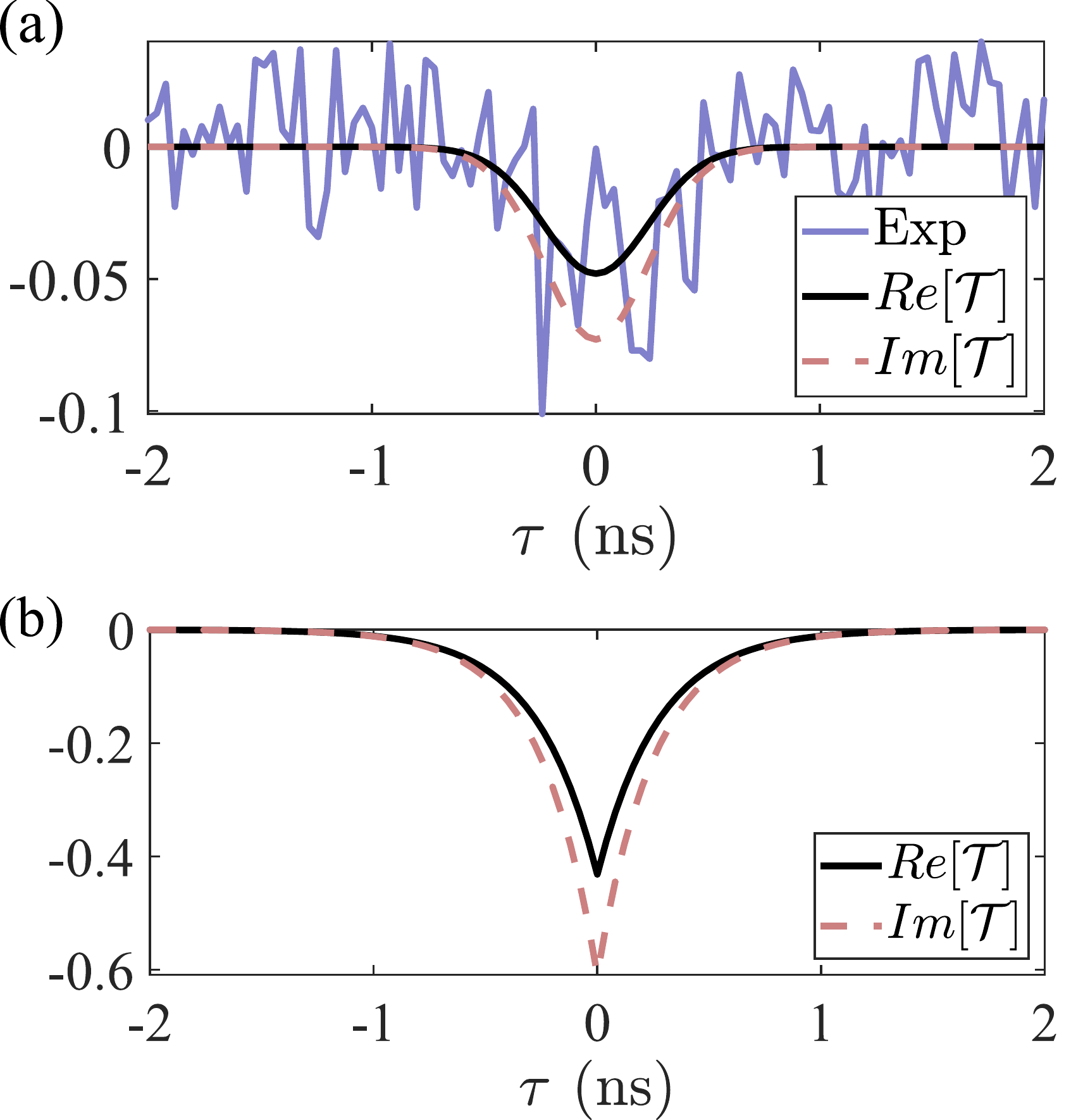}
\caption{(a) Real part of the two-photon correlated coefficient ${\rm Re}[\mathcal{T}]$ reconstructed from experimental data (blue) and comparison to theory (black). The theoretical imaginary part ${\rm Im}[\mathcal{T}]$ is also included (dashed pink). (b) Theoretical prediction of the real (solid black) and imaginary (dashed pink) parts of $\mathcal{T}$ in the absence of external experimental imperfections.}
\label{Figure4}
\end{figure}

As a first step, we use Eq.~(\ref{eq:It}) and the KK relation (adapted for experimental imperfections, see SM \cite{SM}) to extract the single-photon transmission and reflection coefficients from the intensity data $I_t(\omega)$. We plot the experimental amplitude and phase of both $t(\omega)$ and $r(\omega)$ in Fig.~\ref{Figure3}, and observe excellent agreement with theory for the experimentally determined parameters. The asymmetry of the resonances, a minor frequency shift, and the non-zero phase shift away from the resonance are again here due to the Fano effect, cf.~SM for details \cite{SM}. We record an experimental maximal single-photon phase shift of $\approx 150^{\circ}$ in reflection and $\approx -40^{\circ}$ in transmission.

We can now extract the real part of the intrinsic two-photon scattering coefficient $\mathcal{T}(\omega,\tau)$ based on Eq.~(\ref{TwoPhotonExtraction}), $\chi^{\mu}(\omega)$, and the photo-correlation measurements in all directions $g_{\mu\mu'}^{(2)}(\tau)$. Special care was taken to suppress residual stray scattering from the excitation laser for the measurements in reflection to improve the signal-to-noise ratio. The excitation power was a factor of $\approx 3$ higher than in Fig.~\ref{Figure2}(b).
The data are plotted in Fig. \ref{Figure3}(b), displaying bunching in $g^{(2)}_{tt}$, $g^{(2)}_{rt}$, and anti-bunching in $g^{(2)}_{rr}$. We find excellent agreement between experiment and theory using the system parameters and the modeling of imperfections discussed above, cf.~SM \cite{SM} for details.

The experimental reconstruction of the real part of the intrinsic two-photon correlations $\mathcal{T}$, obtained by processing the experimental data, is shown in Fig.~\ref{Figure4}(a). The reconstructed lineshape and depth is found to be in accordance with the theoretical prediction for a TLE, ${\cal T}(\omega,\tau)=-r(\omega)^2e^{-|\tau|(\gamma_{tot}/2-i[\omega-\omega_0])}$, when including all the discussed imperfections [see black line in Fig.~\ref{Figure4}(a)]. This is to our knowledge the first experimental reconstruction of the two-photon nonlinear response. This analysis pinpoints the genuine strength of correlated two-photon response of the QD-waveguide system, and can be extended further to extract the full matrix in Eq. (\ref{eq:T}) straightforwardly by scanning the input laser frequency $\omega$ and applying the KK relation \cite{SM}. This requires high frequency resolution and a broad scanning interval $\sim 5\gamma_{tot}$ to accurately determine the amplitude and phase \cite{Meissner_2012}.

The signal-to-noise ratio in Fig.~\ref{Figure4}(a) is limited by the photon collection efficiency, especially in reflection. Nevertheless, this can be substantially improved by designing a waveguide with three different coupling gratings, so that the excitation and reflected signals are spatially separated (cf.~Sec.~I.F. of SM~\cite{SM}). Another important experimental challenge is to further improve the electrical noise performance of the device so that spectral diffusion of the QD can be strongly suppressed. Progress on this direction has been very recently obtained \cite{Pedersen_2020}. Alternatively, a feedback loop could be implemented to adjust for the slow frequency drift \cite{Hansom_2014}. When fully correcting for these effects, we predict an order of magnitude enhancement of the two-photon nonlinearity induced by the QD [cf.~Fig.~\ref{Figure4}(b)], which would lead to a higher signal-to-noise ratio in the reconstruction. Moreover, we also predict an enhancement by two orders of magnitude in the bunching of $g_{tt}^{(2)}$ and $g_{tr}^{(2)}$ (see Fig.~S2 of SM~\cite{SM}), which shows the capabilities of the highly coherent light-matter interface.

In summary, we have presented measurements of the one- and two-photon components of the scattering matrix of a single QD in a photonic-crystal waveguide excited by a weak laser source. The applied method relies on intensity and second-order photon correlation measurements making it well suited for current experimental settings and devices. Specifically, we have presented the first experimental reconstruction of the intrinsic two-photon scattering correlations that are induced by the appearance of the two-photon bound state \cite{Shen_2007multi,Shen_2010}. Extending this approach to three- or even $N$-photon scattering processes requires measuring $N$-order photon correlations in all $2^N$ possibilities of propagation direction, but this will be discussed elsewhere \cite{Ramos_2020}.

This type of reconstruction technique will enable further developments within quantum nonlinear optics, where a thorough understanding of the nonlinear response is required in potential applications of the nanophotonic hardware. For instance, it has been shown that the two-photon scattering processes, if properly controlled via the incoming photon pulse lengths, can be the basis of deterministic photon sorting, which enables the construction of deterministic Bell analyzers and photonic gates \cite{Witthaut_2012,Ralph_2015}. Furthermore, the presence of exotic photon bound states provides a route to study complex many-body quantum physics \cite{Shen_2010,Mahmoodian_2019}.

\begin{acknowledgements}
The authors would like to thank A. S. S\o{}rensen for fruitful discussions.
We gratefully acknowledge financial support from Danmarks Grundforskningsfond (DNRF 139, Hy-Q Center for Hybrid
Quantum Networks), H2020 European Research Council (ERC) (SCALE), Styrelsen for Forskning og Innovation
(FI) (5072-00016B QUANTECH), Bundesministerium fur Bildung und Forschung (BMBF) (16KIS0867, Q.Link.X),
Deutsche Forschungsgemeinschaft (DFG) (TRR 160). T.R. and J.J.G.-R. acknowledge support from project PGC2018-094792-B-I00 (MCIU/AEI/FEDER, UE), CAM/FEDER project No. S2018/TCS-4342 (QUITEMAD-CM) and CSIC Quantum Technology Platform PT-001. T.R. further acknowledges funding from the EU Horizon 2020 program under the Marie Sk\l{}odowska-Curie grant agreement No. 798397.
\end{acknowledgements}

\beginsupplement
\newpage
\onecolumngrid
\newpage
{
\center \bf \large
Supplemental Material for: \\
Experimental reconstruction of the few-photon nonlinear scattering matrix from a single quantum dot in a nanophotonic waveguide\vspace*{0.1cm}\\
\vspace*{0.0cm}
}

\begin{center}
Hanna Le Jeannic$^{1}$, Tom\'as Ramos$^{2}$, Signe F. Simonsen$^{1}$, Tommaso Pregnolato$^{1}$, Zhe Liu$^{1}$, R\"udiger Schott$^{3}$, Andreas D. Wieck$^{3}$, Arne Ludwig$^{3}$, Nir Rotenberg$^{1}$, Juan Jos\'e Garc\'ia-Ripoll$^{2}$, and Peter Lodahl$^{1}$\\
\vspace*{0.15cm}
\small{\textit{$^1$Center for Hybrid Quantum Networks (Hy-Q), Niels Bohr Institute,\\ University of Copenhagen, Blegdamsvej 17, DK-2100 Copenhagen, Denmark\\
$^2$Instituto de F\'{i}sica Fundamental IFF-CSIC, Calle Serrano 113b, Madrid 28006, Spain\\
$^3$Lehrstuhl f{\"u}r Angewandte Festk{\"o}rperphysik, Ruhr-Universit{\"a}t, Universit{\"a}tsstrasse 150, D-44780 Bochum, Germany}}\\
\vspace*{0.25cm}
\end{center}

\twocolumngrid

\section*{Contents}
\begin{itemize}
\item I.---Realistic modeling of the experimental setup and measurements.
\item II.---Derivation of few-photon scattering matrices and reconstruction relations.
\end{itemize}

\section{I.- Realistic modeling of the experimental setup and measurements}

In this section we provide the theory to describe the scattering experiment of a quantum dot (QD) interacting with a coherent state input and coupled to a photonic crystal waveguide (PhCW), including the effect of various experimental imperfections.

In Sec.~I.A, we state the quantum Langevin equations, input-output relations, and master equation describing our system in the presence of a Fano resonance and white noise dephasing. In Sec.~I.B and Sec.~I.C, we give formulas to calculate the intensity $I_\mu(t)$ and second-order correlations functions $g_{\mu\mu'}^{(2)}(\tau)$ predicted by the model. In Sec.~I.D, we show how to include imperfections from spectral diffusion, finite instrument response time, and background noise. In Sec.~I.E, we extract the experimental parameters of the setup by fitting the full model to the experimental data from intensity and second-order correlation measurements. Finally, in Sec.~I.F, we give details on the PhCW and the laser setup used in our experiments.

\subsection{I.A.- Dynamics of the system including radiative loss, dephasing, and Fano resonance}

We model the QD as a two-level emitter (TLE) of transition frequency $\omega_0$ and we consider its interaction with a coherent state $|\alpha\rangle_t$ through the transmission port of the waveguide. The initial state $|\Psi_{\rm in}\rangle$ of the total system is then given by 
\begin{align}
|\Psi_{\rm in}\rangle=|\alpha\rangle_t|0\rangle_r|0\rangle_l|\Psi_{\rm TLE}\rangle,\label{InitialState}
\end{align}
where $|\Psi_{\rm TLE}\rangle$ is a generic initial state of the TLE, and $|0\rangle_r,|0\rangle_l$ are photonic vacuum states in the reflection port and loss channels, respectively. The coherent state input on the transmission channel is defined as $|\alpha\rangle_t=e^{-|\alpha|^2/2}e^{\alpha A^\dag}|0\rangle_t$, where $A^\dag=\int d\omega f_{\bar{\omega}}(\omega)a_{\rm in}^t{}^\dag(\omega)$ creates a photon in a wavepacket with profile $f_{\bar{\omega}}(\omega)$, centered at $\bar{\omega}$, and $a_{\rm in}^{\mu}(\omega)$ is the standard annihilation operator of a monochromatic input photon of frequency $\omega$ in direction $\mu=t,r$.

The quantum Langevin equations [SM1] describing the dynamics of the TLE in the presence of a Fano resonance and white noise pure dephasing have been previously studied [SM2, SM3], and can be written in a rotating frame with the wavepacket central frequency $\bar{\omega}$ as
\begin{align}
\frac{d \sigma^{-}}{dt}={}&-\left(\frac{\gamma_{tot}}{2}+\gamma_d-i[\bar{\omega}-\omega_0]\right)\sigma^{-}+i\sigma_z\sqrt{\gamma_{l}}b_{\rm in}(t)\nonumber\\
{}&+i\sigma_z\sum_{\mu=t,r}\sqrt{\frac{\beta\gamma_{tot}}{2}}a^{\mu}_{\rm in}(t)-i\sqrt{2\gamma_d}\Theta(t)\sigma^-,\label{EqMotSm}\\
\frac{d\sigma_z}{dt}={}&-\gamma_{tot}(\sigma_z+1)-2i\sqrt{\gamma_{l}}(\sigma^{+}b_{\rm in}(t)-{\rm h.c.})\nonumber\\
{}&-2i\sum_{\mu=t,r}\sqrt{\frac{\beta\gamma_{tot}}{2}}\left(\sigma^{+}a_{\rm in}^{\mu}(t)-{\rm h.c.}\right).\label{EqMotSz}
\end{align}
Here, $\sigma^-(t)$, $\sigma^+(t)$, and $\sigma_z(t)=2\sigma^+\sigma^--1$ are the standard lowering, raising, and population difference operators of the TLE. In addition, $\beta=\gamma_{wg}/\gamma_{tot}$ is the coupling efficiency, with $\gamma_{wg}$ the decay into the waveguide, $\gamma_{tot}=\gamma_{wg}+\gamma_l$ the total radiative decay, and $\gamma_l$ the unwanted loss into other modes. The dephasing rate is denoted by $\gamma_d$ and $\Theta(t)$ is the classical stochastic field causing the dephasing noise, which is delta-correlated for white noise, i.e.~$\langle\!\langle \Theta(t)\Theta(t') \rangle\!\rangle=\delta(t-t')$ [SM2]. The input noise operator for photons going to the loss channels is denoted by $b_{\rm in}(t)$. Most importantly for our scattering experiments, $a_{\rm in}^{\mu}(t) = (2\pi)^{-1/2}\int d\omega e^{-i(\omega-\bar{\omega}) t} a_{\rm in}^{\mu}(\omega)$ is the input field of photons entering the waveguide through the port $\mu=t,r$, and these are related to the output field operators by the input-output relations [SM2, SM3],
\begin{align}
a_{\rm out}^{\mu}(t)=\sum_{\mu'=t,r}\Lambda_{\mu\mu'}a_{\rm in}^{\mu'}(t)-i\frac{z^2}{|z|^2}\sqrt{\frac{\beta\gamma_{tot}}{2}}\sigma^-(t).\label{InputOutputEq}
\end{align}
Here, the coefficients $\Lambda_{\mu\mu'}=\delta_{\mu\mu'}+z-1$, and
\begin{align}
z=\frac{1}{1+i\xi},
\end{align}
characterize the internal reflections due to the Fano resonance. These reflections can be modeled by coupling the QD (the "discrete state") to a lossy and off-resonant parasitic cavity mode (the "continuum of states"), which can be adiabatically eliminated [SM2, SM3], obtaining $\xi=\Delta_c/\kappa$ with $\kappa$ the cavity decay rate and $\Delta_c$ the cavity detuning with respect to the QD. Throughout our work, $\xi$ will be approximated to be constant over the small frequency range considered around the QD resonance.

When numerically calculating measurable expectation values and correlations functions with the present model, it will be more convenient to use a master equation formalism for the TLE dynamics, and then connect to the photonic operators via the input-output relations (\ref{InputOutputEq}). Indeed, the master equation that is equivalent to the quantum Langevin equations (\ref{EqMotSm})-(\ref{EqMotSz}) reads
\begin{align}
	\dot{\rho}=-i[H,\rho]+\gamma_{tot} {\cal D}[\sigma^-]\rho+2\gamma_d{\cal D}[\sigma^+\sigma^-]\rho,\label{MasterEq}
\end{align}
with $\rho(t)$ the mixed state of the TLE, ${\cal D}[x]\rho=x\rho x^\dag -(x^\dag x \rho - \rho x^\dag x)/2$ the Lindblad operator [SM1], and $H$ the Hamiltonian the driven TLE,
\begin{align}
H(t)=-(\bar{\omega}-\omega_0)\sigma^+\sigma^- + i\Omega(t)(\sigma_+-\sigma_-).    
\end{align}
Here, $\Omega(t) = -i\alpha f(t) \sqrt{\beta\gamma_{tot}/2}$ is the effective coherent driving strength on the TLE, whose time-dependence $f(t)$ is given by the Fourier transform of the wavepacket profile $f(t)=(2\pi)^{-1/2}\int d\omega e^{-i(\omega-\bar{\omega}) t} f_{\bar{\omega}}(\omega)$.

In the experiment presented in this work, we shine the TLE with a monochromatic cw laser with frequency $\bar{\omega}=\omega$ and constant photon flux $F$. In this special case, the wavepacket profile can be represented as [SM4]
\begin{align}
    f_{\omega}(\omega')=\sqrt{2\pi F}\delta(\omega'-\omega),\label{cwlaser}
\end{align}
and the driving strength $\Omega = -i\alpha\sqrt{\beta F \gamma_{tot}/2}$ is stationary. This implies that for a cw laser input, the weak driving condition $|\Omega|^2/\gamma_{tot}^2\ll 1$ is equivalent to $|\alpha|^2 (F/\gamma_{tot})(\beta/2)\ll 1$.

\subsection{I.B.- Intensity measurements}

Experimentally, we shine a cw laser (\ref{cwlaser}) on the quantum dot and measure in steady state (ss) the output intensity of light $I_\mu(\omega)$ by recording the counts of a single-photon detector (SPD) at each output port of the waveguide $\mu=t,r$. Theoretically, these measurements are described by
\begin{align}
I_{\mu} = \frac{\langle a_{\rm out}^{\mu}{}^\dag(t) a_{\rm out}^{\mu}(t) \rangle_{\rm ss}}{ |z|^2 |\alpha|^2 F},\label{Ideff}
\end{align}
where the normalization by $|z|^2$ and the input photon flux $|\alpha|^2F$ is chosen so that the intensity at transmission becomes unity ($I_t(\omega)\rightarrow 1$) off resonance $(|\omega-\omega_0|\rightarrow \infty)$.

To relate the photon intensity (\ref{Ideff}) to expectation values of the TLE, we use the input-output relation (\ref{InputOutputEq}) and take expectation values on the input state (\ref{InitialState}), obtaining
\begin{align}
I_{\mu} ={}&\frac{|\Lambda_{\mu t}|^2}{|z|^2}+\frac{\beta\gamma_{tot}}{|z|^2}{\rm Re}\left[\left(\frac{\beta}{2}-\frac{z^2}{|z|^2}\Lambda_{\mu t}^\ast\right)\frac{\langle \sigma^-\rangle_{\rm ss}}{\Omega}\right].\label{IntensityExplicit}
\end{align}
Here, $\langle \sigma^- \rangle_{\rm ss}$ is the coherence of the TLE at steady state ($t\rightarrow \infty$), which is calculated from steady state solution $\rho_{\rm ss}$ of the master equation (\ref{MasterEq}) as $\langle \sigma^- \rangle_{\rm ss}={\rm Tr}\lbrace \sigma^- \rho_{\rm ss}\rbrace$. When fitting the intensity measurements to our model (cf.~Sec~I.E), we numerically calculate the expression (\ref{IntensityExplicit}) as a function of the monochromatic laser frequency $\omega$ and driving intensity $\Omega$, and then we average the result over the remaining noise sources treated in Sec.~I.D. Notice that to derive Eq.~(\ref{IntensityExplicit}), we have used the properties $a_{\rm in}^\mu(t)|\Psi_{\rm in}\rangle=\alpha \sqrt{F}\delta_{\mu t}|\Psi_{\rm in}\rangle$ and $\langle \sigma^+ \sigma^- \rangle_{\rm ss} = 2 {\rm Re}[\frac{\Omega^\ast}{\gamma_{tot}}\langle\sigma^-\rangle_{\rm ss}]$, obtained by formally integrating Eq.~(\ref{EqMotSz}) [SM2].

\subsection{I.C.- Second-order photo-correlation measurements}

The key requirement to perform the two-photon scattering reconstruction is the ability to measure second-order correlation functions $ g^{(2)}_{\mu\mu'}(\tau)$, at all combination of output ports $\mu,\mu'=t,r$, and with a time delay $\tau$ between the clicks of two SPDs.

In the case of a cw laser input, the $g^{(2)}_{\mu\mu'}(\tau)$ correlations at channels $\mu,\mu'=t,r$ can be expressed in terms of the output photon operators as
\begin{align}
g^{(2)}_{\mu\mu'}(\tau) = \frac{G^{(2)}_{\mu\mu'}(\tau)}{I_\mu I_{\mu'}},\label{g2def}
\end{align}
where $I_\mu$ are the intensities in Eq.~(\ref{IntensityExplicit}), and $G^{(2)}_{\mu\mu'}(\tau)$ are the unnormalized two-photon correlation functions,
\begin{align}
G^{(2)}_{\mu\mu'}(\tau) = \frac{\langle a_{\rm out}^{\mu}{}^\dag(t) a_{\rm out}^{\mu'}{}^\dag(t+\tau) a_{\rm out}^{\mu'}(t+\tau) a_{\rm out}^{\mu}(t) \rangle_{\rm ss}}{|z|^4|\alpha|^4F^2}.\label{G2def}
\end{align}

Analogously to $I_\mu(\omega)$ in Sec.~I.B, the two-photon correlations $G^{(2)}_{\mu\mu'}(\tau)$ can be related to the TLE steady state expectation values and correlations using the input-output relations (\ref{InputOutputEq}), and the properties $a_{\rm in}^\mu(t)|\Psi_{\rm in}\rangle=\alpha \sqrt{F}\delta_{\mu t}|\Psi_{\rm in}\rangle$, and $[\sigma^-(t),a_{\rm in}^\mu(t')]=[\sigma^+(t),a_{\rm in}^\mu(t')]=0$, $t\geq t'$, obtaining
\begin{align}
G^{(2)}_{\mu\mu'}(\tau) ={}& \frac{|\Lambda_{\mu t}|^2|\Lambda_{\mu' t}|^2}{|z|^4}\label{G2Explicit}\\
-{}& \frac{\beta\gamma_{tot}}{|z|^4} {\rm Re}\left[\left(|\Lambda_{\mu t}|^2 \Lambda_{\mu't}^\ast+|\Lambda_{\mu' t}|^2 \Lambda_{\mu t}^\ast\right)\frac{z^2}{|z|^2}\frac{\braket{\sigma^-}_{\rm ss}}{\Omega}\right]\nonumber\\
+{}&\frac{\beta^2\gamma_{tot}^2}{4|z|^4}\left(|\Lambda_{\mu t}|^2+|\Lambda_{\mu' t}|^2\right)\frac{\braket{\sigma^+\sigma^-}_{\rm ss}}{|\Omega|^2}\nonumber\\
+{}&\frac{\beta^2\gamma_{tot}^2}{2|z|^4}{\rm Re}\left[\Lambda_{\mu t}^\ast \Lambda_{\mu't}^\ast\frac{z^4}{|z|^4}\frac{\braket{\sigma^-(t+\tau)\sigma^-(t)}_{\rm ss}}{\Omega^2}\right]\nonumber\\
+{}&\frac{\beta^2\gamma_{tot}^2}{2|z|^4}{\rm Re}\left[\Lambda_{\mu t}^\ast \Lambda_{\mu't}\frac{\braket{\sigma^+(t+\tau)\sigma^-(t)}_{\rm ss}}{|\Omega|^2}\right]\nonumber\\
-{}&\frac{\beta^3\gamma_{tot}^3}{4|z|^4}{\rm Re}\left[ \Lambda_{\mu't}^\ast\frac{z^2}{|z|^2}\frac{\braket{\sigma^+(t)\sigma^-(t+\tau)\sigma^-(t)}_{\rm ss}}{\Omega|\Omega|^2}\right]\nonumber\\
-{}&\frac{\beta^3\gamma_{tot}^3}{4|z|^4}{\rm Re}\left[ \Lambda_{\mu t}^\ast\frac{z^2}{|z|^2}\frac{\braket{\sigma^+(t+\tau)\sigma^-(t+\tau)\sigma^-(t)}_{\rm ss}}{\Omega|\Omega|^2}\right]\nonumber\\
+{}&\frac{\beta^4\gamma_{tot}^4}{16|z|^4}\frac{\braket{\sigma^+(t)\sigma^+(t+\tau)\sigma^-(t+\tau)\sigma^-(t)}_{\rm ss}}{|\Omega|^4}\nonumber.
\end{align}

Using the master equation (\ref{MasterEq}) and the quantum fluctuation-regression theorem [SM1], we can calculate all the TLE expectation values and the two-time correlation functions appearing in Eq.~(\ref{G2Explicit}) and thereby obtain $G^{(2)}_{\mu\mu'}(\tau)$. Then, normalizing by the intensity $I_\mu$ in Eq.~(\ref{IntensityExplicit}), we get the photon correlation function $g^{(2)}_{\mu\mu'}(\tau)$ using Eq.~(\ref{g2def}). This calculation is valid for general driving strength $\Omega$, laser frequency $\omega$, Fano resonance $\xi$, and white noise dephasing rate $\gamma_d$. In the next Sec.~I.D, we show how to include other imperfections to this result and thereby obtain the final expression for $g^{(2)}_{\mu\mu'}(\tau)$ that we use to fit our  measurements [cf.~Sec.I.E].

\subsection{I.D.- Spectral diffusion, finite instrument response function, and background noise}

In this section, we show how to include the final three imperfections we observe in our experiment: spectral diffusion due to a slow frequency drift of the QD, instrument response function (IRF), due to a finite time window of the SPDs, background noise on the waveguide channels.

First, we consider the spectral diffusion effect, which we model by averaging the intensity $I_t$ and the correlation functions $g^{(2)}_{\mu\mu'} (\omega,\tau)$ with a Gaussian distribution of deviations $\Delta$ from the average QD resonance frequency $\omega_0$. Since this effect can be manifested differently depending on the time scales of the measurements, we use two different standard deviations: $\sigma_{\mathrm{short}}$ for $I_\mu(\omega)$ measurements whose acquisition time is on order of a minute, and $\sigma_{\mathrm{long}}$ for the photon correlation measurements $g^{(2)}_{\mu\mu'}$ as their acquisition time is on order of hours. Applying this averaging to the intensity and second-order correlation functions, we obtain
\begin{align}
\bar{I}_\mu={}&\int d\Delta P_{\mathrm{SD}}(\Delta,\sigma_{\mathrm{short}}) I_{\mu}(\omega-\Delta),\\
\bar{g}_{\mu\mu'}^{(2)}(\omega,\tau)={}&\frac{\int d\Delta P_{\mathrm{SD}}(\Delta,\sigma_{\mathrm{long}}) G^{(2)}_{\mu\mu'}(\omega-\Delta,\tau)}{\int d\Delta P_{\mathrm{SD}}(\Delta,\sigma_{\mathrm{long}})I_{\mu}(\omega-\Delta)I_{\mu'}(\omega-\Delta)},
\end{align}
where the Gaussian distribution is defined as
\begin{align}
P_{\mathrm{SD}}(\Delta,\sigma)=\frac{1}{\sqrt{2\pi\sigma^2}} e^{-\frac{\Delta^2}{2\sigma^2}},\label{GaussSD}
\end{align}
with $\sigma_{\mathrm{long}}$ and $\sigma_{\mathrm{short}}$ corresponding to the long and short timescales for spectral diffusion.

The second imperfection is the IRF. Due to the finite response window of our SPDs, the observed linshape will be broader than predicted by $\bar{g}^{(2)}_{ \mu \mu'} (\omega,\tau)$. We model this extra broadening observed in experiments by convolving the $\bar{g}^{(2)}_{ \mu \mu'} (\omega,\tau)$ with another Gaussian distribution $P^{\mathrm{IRF}}(\tau')$ describing the IRF. Similarly to the spectral diffusion, we obtain
\begin{equation}
\tilde{g}_{\mu\mu'}^{(2)}(\omega,\tau)=\int d\tau' P_{\mathrm{IRF}}(\tau'-\tau) \bar{g}_{\mu\mu'}^{(2)}(\omega,\tau').
\end{equation}
Here, the Gaussian distribution of time delays
\begin{equation}
P_{\mathrm{IRF}}(\tau')=\frac{1}{\sqrt{2\pi\sigma_{\mathrm{IRF}}^2}} e^{-\frac{\tau'^2}{2\sigma_{\mathrm{IRF}}^2}},\label{GaussIRF}
\end{equation}
has a standard deviation of $\sigma_{\mathrm{IRF}}=200$ ps, which was determined independently.

Finally, we also consider the possibility for background noise $B_{\mu\mu'}$ appearing on the different channels $\mu,\mu'=t,r$, and which modify the measured second-order correlation functions as
\begin{align}
    \hat{g}^{(2)}_{\mu\mu'}(\tau)=\frac{(1-B_{\mu\mu'})G^{(2)}_{\mu\mu'}(\tau)+B_{\mu\mu'}\times 1}{(1-B_{\mu\mu'})I_{\mu}I_{\mu'}+B_{\mu\mu'}\times 1}.
\end{align}
This background will only appear in $\hat{g}^{(2)}_{rr}(\tau)$. It originates from imperfect extinction of the incoming laser light.

\subsection{I.E.- Extraction of parameters and errors}

In this section, we numerically calculate our full theoretical predictions for intensity and correlation functions including all the imperfections discussed above, and we fit them to our experimental data of $I_t$ and $g^{(2)}_{\mu\mu'} (\omega,\tau)$, shown in Fig.~S2 and Fig.~S4.(a) of the main text, as well as in Fig.~S5 of this SM. From this analysis we extract all the relevant system parameters of our setup, which are displayed in Table~\ref{tab:TableParam}.

Before performing the fits, we independently measure the total radiative decay rate $\gamma_{tot}$ of the QD. To do so, we first send a 9-ps laser pulse via the waveguide and on resonance with the QD. The time of emission of the photons is aquired with a 20-ps jitter avalanche photodiode (APD), registered with a time tagger, and finally fitted with a single exponential decay convolved with the IRF of the APD. This results in a fitted lifetime of 131 ps, corresponding to a total QD decay rate of $\gamma_{tot}/2\pi=(\gamma_{wg}+\gamma_{l})/2\pi=1.22$~GHz. This measurement results are shown in the inset of Fig.~2(a).

\begin{figure}[t!]
\centering
\includegraphics[width=8.6 cm]{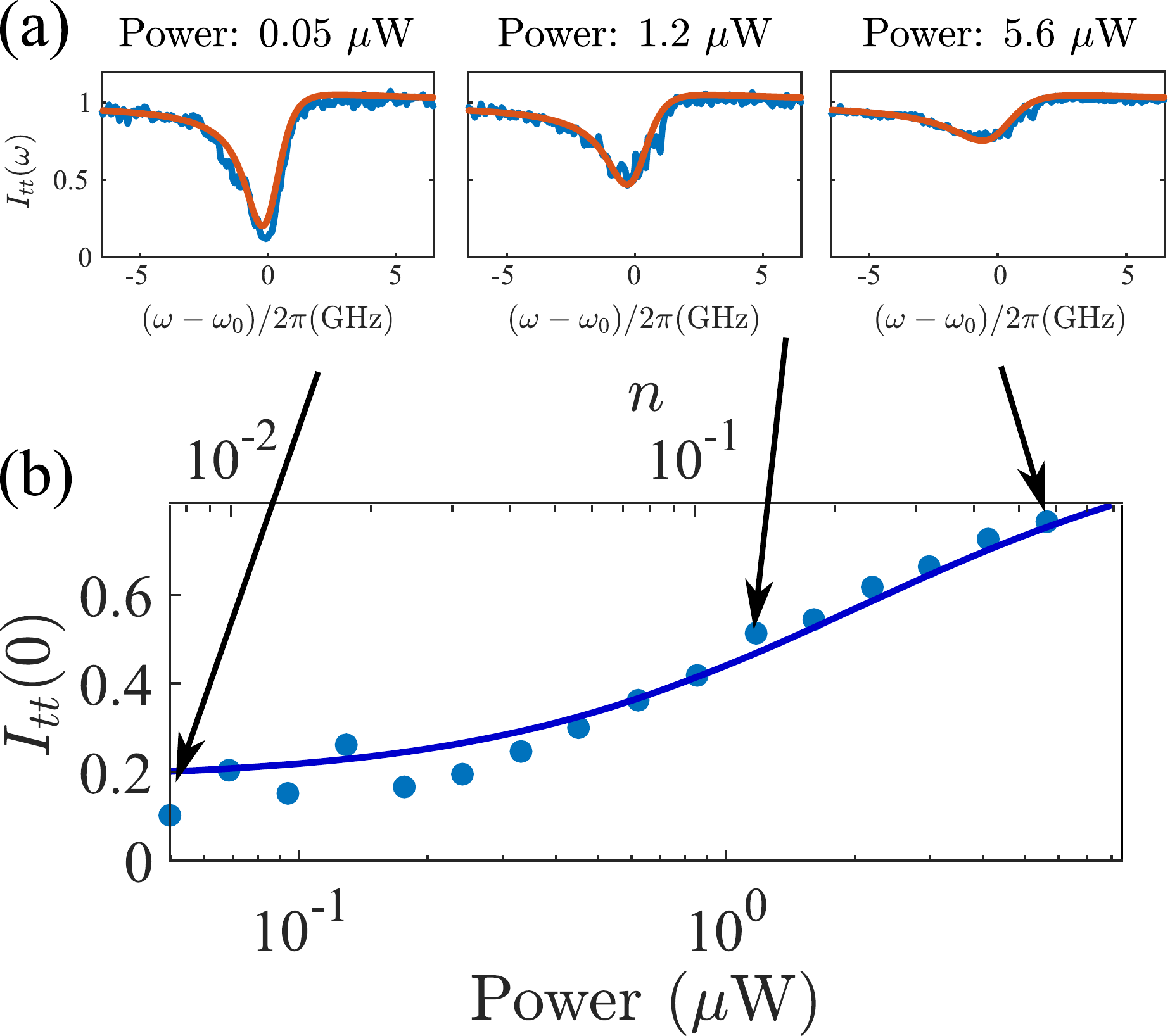}
\caption{(a)Measurement of the transmission intensity  $I_t$ as a function of laser frequency $\omega$ for different input powers. The actual measurements are shown in blue, while the theoretical prediction for the extracted system parameters is shown in red. (b) Saturation curve and the corresponding fit of $I_t$ as a function of laser power (lower horizontal axis) and as a function of mean photon number per lifetime $n=2\frac{\Omega^2}{\gamma_{tot}^2}$.}
\refstepcounter{SIfig}\label{SupFigfit}
\end{figure}

In addition, we probe the behavior of the transmitted intensity of the QD as a function of laser power, which displays the characteristic saturation curve as a function of the mean photon number per lifetime $n=2\frac{\Omega^2}{\gamma_{tot}^2}$. [cf.~Fig.~\ref{SupFigfit}]. We use these measurements, in addition to the second-order correlation measurements $g^{(2)}_{\mu\mu'} (\omega,\tau)$, shown in Fig.~3 of the main text, to fit the full theoretical model and thereby extract the experimental parameters describing the system. We perform on this purpose a least square fit on the full dataset at hand. Since $\gamma_{tot}$ is independently measured, from this analysis we obtain the Fano parameter $\xi$, the loss factor $\eta$ of the setup relating the input measured laser power to the Rabi frequency in the waveguide mode $\Omega$ ($\Omega= \sqrt{\eta P}$), the background noise coefficients $B_{\mu\mu'}$, the spectral diffusion standard deviations $\sigma_{\mathrm{short}}$, $\sigma_{\mathrm{long}}$, the coupling efficiency $\beta$, and the white noise dehasing $\gamma_d$. All these experimentally extracted parameters and respective $95\%$ confidence intervals at are presented in Table \ref{tab:TableParam}.

\begin{table}[ht]
\centering
\caption{Parameters extracted from fitting the experimental data sets and their error ranges.}
\begin{tabular}[t]{c  c  c }
\hline \hline
    Parameter & Value & Confidence interval ($95\%$) \\ \hline
    $\beta$  & $0.87$ & $\left[0.83,0.91\right]$ \\
    $\gamma_{d}$ & $\simeq 0 \mathrm{ns}^{-1}$ & $ \left[ 0,0.12 \right]$ \\
    $\sigma_{\mathrm{short}}$  & $330 \mathrm{MHz}$ & $\left[290,360\right]$ \\
    $\sigma_{\mathrm{long}}$  & $660 \mathrm{MHz}$ & $\left[500, 820\right]$ \\
	$\xi$  & $-0.26$ & $\left[-0.27, -0.25\right]$\\
	$\eta$ & $0.11$ & $\left[0.10,0.12\right]$\\
	$ B_{RR}$ &  $0.07$ & $\left[0.04,0.11\right]$\\
	\hline \hline
\end{tabular}
\label{tab:TableParam}
\end{table}

\begin{figure}[!t]
\includegraphics[width=8.6 cm]{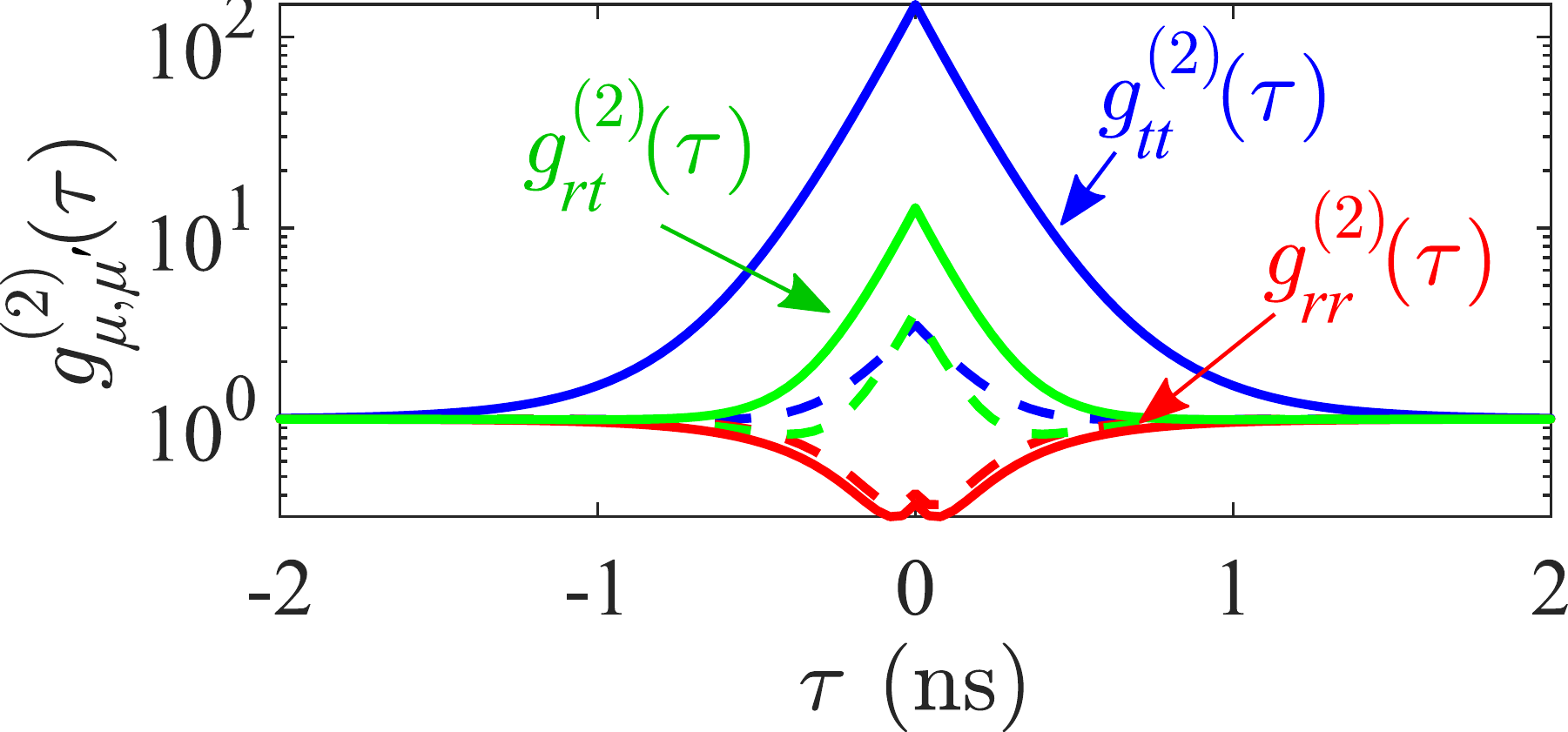}
\caption{Second-order correlation functions $g_{tt}^{(2)}(\tau)$ (blue), $g_{rr}^{(2)}(\tau)$ (red), and $g_{tr}^{(2)}(\tau)$ (green), predicted for our system after correcting the laser background and finite time resolution of the detectors (dashed lines), and additionally in the absence of spectral diffusion (solid lines)}
\label{SupFigIdeal1}
\end{figure}

The resulting fits can be seen on Fig.~3, and Fig.~\ref{SupFigfit}.

In Fig.~\ref{SupFigIdeal1}, we display predictions for $g_{tt}^{(2)}(\tau)$ (blue), $g_{rr}^{(2)}(\tau)$ (red), and $g_{tr}^{(2)}(\tau)$ (green), that our system can have in the future if we correct certain imperfections discussed above. In particular,  the dashed lines correspond to the case where the laser background and finite time resolution of the detectors is corrected, and the solid lines correspond to the case where in addition, spectral diffusion is also corrected. This shows the highly coherent quantum behaviour that  may be achievable in our system.

\subsection{I.F.- The waveguide sample and laser setup}

\begin{figure}[t]
\centering
\includegraphics[width=8.6 cm]{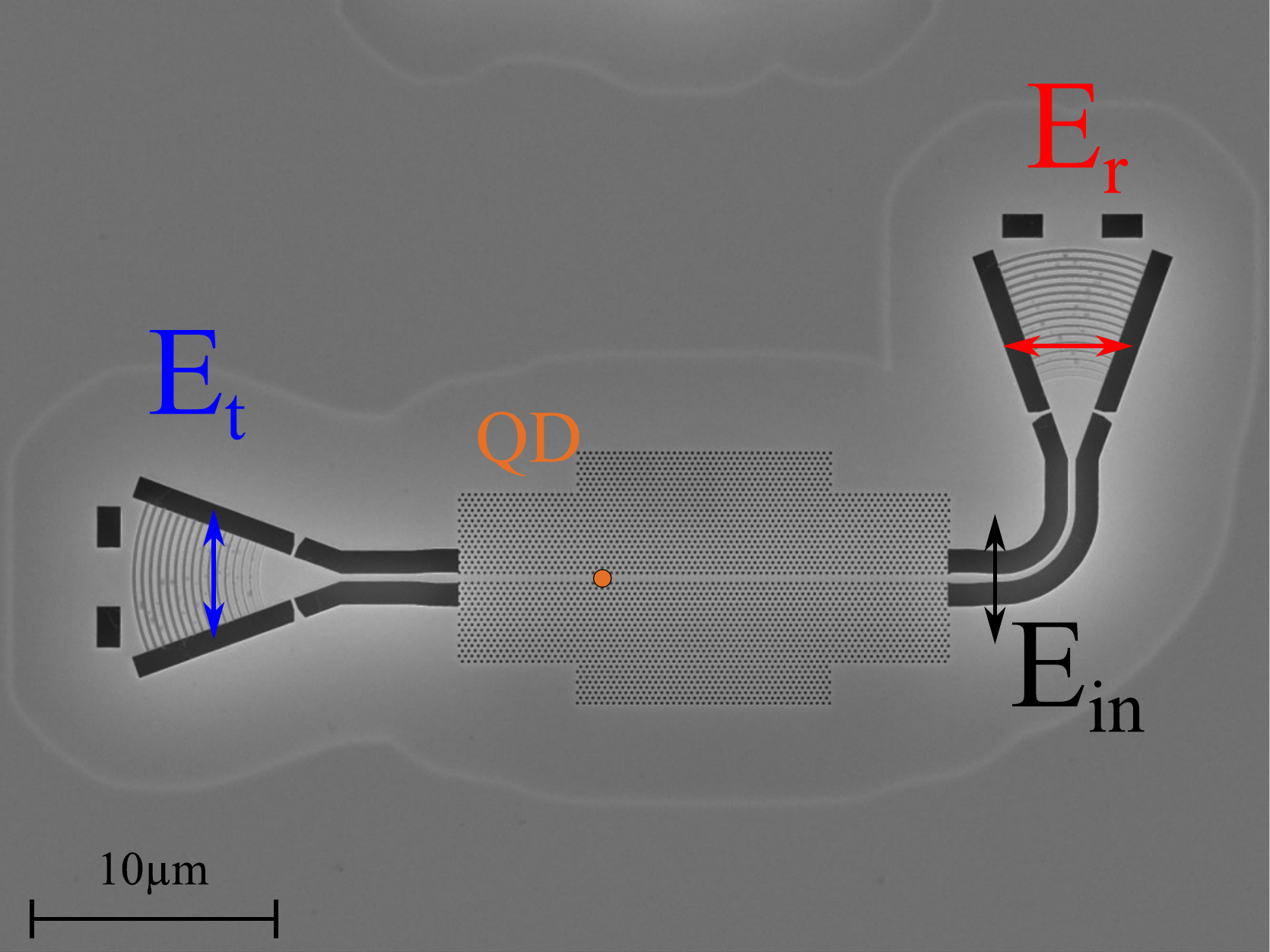}
\caption{SEM picture of the waveguide and configuration and polarizations of the excitation ($E_{in}$) and collection ($E_{t}$ and $E_{r}$) beams.}
\label{SupFigureSEM}
\end{figure}

In order to collect both the reflected and transmitted light we implement a special setup where the incoming laser light is coupled to the photonic crystal waveguide through the nanobeam waveguide section connecting the PhCW to the reflection grating, as depicted in Fig.~\ref{SupFigureSEM}.
We align the incident laser beam in such a way that it is focused relatively far from the reflection-grating coupler and at an angle with respect to the surface of the sample.
The incoming laser light is additionally sent in with a polarization orthogonal to the one optimal for the grating coupler to ensure a higher extinction.
By optimizing the alignment and the polarization, we obtain a resulting extinction of $\simeq 20$ between the laser and the reflected collected light. This finite extinction is taken into account in the fitting of the parameters as an additional background.
The extinction is highly limited by the setup and the alignment scheme used in the experiment, but it could be drastically increased by an improved design of the waveguide. For example, a third grating coupler could be connected to the PhCW by adding a Y-splitter between the PhCW and the reflection grating. This new geometry would results in a better spatial separation between the excitation and collection of the reflected signal.

\section{II.- Derivation of few-photon scattering matrices and reconstruction relations}

In this section, we derive the reconstruction relations shown in the main text and experimentally implemented in this work. In Sec.~II.A we first give the standard definitions of single- and two-photon scattering matrices. In Sec.~II.B, we recall details on the single-photon reconstruction published in Ref.~[SM2] and experimentally implemented in this work. In Sec.~II.C, we show the derivation of the two-photon scattering matrix reconstruction formulas, and finally in Sec.~II.D we show the theoretically predicted two-photon correlations that we experimentally verify in this work.

\subsection{II.A.- Definition of single- and two-photon scattering matrices}

In the input-output formalism, the single- and two-photon scattering matrices read [SM5],
\begin{align}
    S^{\mu}_{\nu_1\omega_1}={}&\langle g|\langle 0| a_{\rm out}^\mu(\nu_1) a_{\rm in}^{t}{}^\dag(\omega_1)|0\rangle|g\rangle,\label{s1Def1}\\
    S^{\mu\mu'}_{\nu_1\nu_2\omega_1\omega_2}={}&\langle g|\langle 0| a_{\rm out}^\mu(\nu_1)a_{\rm out}^{\mu'}(\nu_2) a_{\rm in}^{t}{}^\dag(\omega_1)a_{\rm in}^{t}{}^\dag(\omega_2)|0\rangle|g\rangle,\label{s2Def1}
\end{align}
which describe the probability amplitude that input photons with frequencies $\omega_1,\omega_2$ propagating along the waveguide in the transmission direction are scattered into output photons with frequencies $\nu_1,\nu_2$ and propagation directions $\mu,\mu'=t,r$. The input and output operators in the frequency domain are related to the ones in Eq.~(\ref{InputOutputEq}) by a standard Fourier transform, i.e.
\begin{align}
    a^{\mu}_{\rm in/out}(\omega)={}&\frac{1}{\sqrt{2\pi}}\int dt e^{i\omega t} a^{\mu}_{\rm in/out}(t).\label{timeFreqRel}
\end{align}

The scattering matrix $U$ relates the asymptotic input and output states as
\begin{align}
|\Psi_{\rm out}\rangle=U|\Psi_{\rm in}\rangle,\label{ScatteringAction}
\end{align}
or alternatively, the input and output operators as
\begin{align}
    a^\mu_{\rm out}(\omega)=U^\dag a^\mu_{\rm in}(\omega) U,\label{OutUIn}
\end{align}
and therefore Eqs.~(\ref{s1Def1})-(\ref{s2Def1}) can also be written only in terms of input operators as
\begin{align}
    S^{\mu}_{\nu_1\omega_1}={}&\langle g|\langle 0| a_{\rm in}^\mu(\nu_1) U a_{\rm in}^{t}{}^\dag(\omega_1)|0\rangle|g\rangle,\label{s1Def2}\\
    S^{\mu\mu'}_{\nu_1\nu_2\omega_1\omega_2}={}&\langle g|\langle 0| a_{\rm in}^\mu(\nu_1)a_{\rm in}^{\mu'}(\nu_2) U a_{\rm in}^{t}{}^\dag(\omega_1)a_{\rm in}^{t}{}^\dag(\omega_2)|0\rangle|g\rangle,\label{s2Def2}
\end{align}
where we have assumed $U|0\rangle = |0\rangle$, i.e.~ that photons cannot be created from vacuum.

Finally, using the conservation of energy, photon number, and the isotropy of the light-matter interaction, one can show that the scattering matrices can be finally recast as [SM5]
\begin{align}
    S^{\mu}_{\nu\omega}={}&\chi^{\mu}(\omega) \delta(\nu-\omega),\label{s1Def212}\\
    S^{\mu\mu'}_{\nu_1\nu_2\omega_1\omega_2}={}&\chi^{\mu}(\nu_1)\chi^{\mu'}(\nu_2)\left(\delta(\nu_1-\omega_1)\delta(\nu_2-\omega_2)+[\omega_1\leftrightarrow\omega_2]\right)\nonumber\\
    {}&+T_{\nu_1\nu_2\omega_1\omega_2}\delta(\nu_1+\nu_2-\omega_1-\omega_2),\label{s2Def22}
\end{align}
where $\chi^\mu(\omega)$ are the single-photon transmission ($\chi^t(\omega)=t(\omega)$) and reflection ($\chi^r(\omega)=r(\omega)$) coefficients and $T_{\nu_1\nu_2\omega_1\omega_2}$ the intrinsic two-photon correlation term. In the following, our task is to relate these quantities to the measurements of intensity $I_\mu$ and photon correlations $g_{\mu\mu'}^{(2)}$ discussed above in Sec.~I.

\subsection{II.B.- Single-photon reconstruction formulas}

Using the definition of the single-photon scattering matrix in Eqs.~(\ref{s1Def1}) and (\ref{s1Def212}), the input-output relation (\ref{InputOutputEq}) and the Langevin equations (\ref{EqMotSm})-(\ref{EqMotSz}), we can show using the standard procedure [SM5] that transmission and reflection coefficients of a TLE with correlated dephasing noise and in the presence of Fano resonance read,
\begin{align}
t(\omega) ={}& z \left(1-\frac{z\beta}{|z|^2} G(\omega)\right),\label{singlePhotonTransmission}\\
r(\omega) ={}& t(\omega)-1.\label{singlePhotonreflection}
\end{align}
Here, $G(\omega)$ is the Laplace transform of the Ramsey dephasing function $C_\phi(t)$, namely $G(\omega)= (\gamma_{tot}/2)\int_{0}^\infty dt e^{-(\gamma_{tot}/2-i[\omega-\omega_0])t} C_\phi (t)$ [cf.~Ref.~SM2]. For instance, in the case of white noise dephasing that we consider in our experiment, we have $C_\phi(t)=e^{-\gamma_d t}$, and therefore
\begin{align}
G(\omega) = \frac{\gamma_{tot}/2}{\gamma_{tot}/2+\gamma_d-i(\omega-\omega_0)}.\label{EqG}
\end{align}

From Eqs.~(\ref{singlePhotonTransmission})-(\ref{singlePhotonreflection}), we see that finding the single-photon scattering coefficients is reduced to experimentally determining $G(\omega)$. To do so, we follow the procedure in Ref.~[SM2] and experimentally extract this quantity from the intensity transmission measurements  $I_t(\omega)$. This requires measuring $I_t$ with an attenuated laser power, i.e.~$|\Omega|^2/\gamma_{tot}^2=|\alpha|^2 (F/\gamma_{tot})(\beta/2)\ll 1$. In this case, it is shown in Ref.~[SM2] that steady state coherence of the TLE can be related to the $G(\omega)$ function we want as
\begin{align}
\frac{\langle \sigma^- \rangle_{\rm ss}}{\Omega} = \frac{G(\omega)}{\gamma_{tot}/2} + {\cal O}\left(|\alpha|^2\right).\label{approxxx}
\end{align}
Using Eq.~(\ref{approxxx}) into the full expression for $I_t(\omega)$ in Eq.~(\ref{IntensityExplicit}), we find the transmission intensity can be conveniently expressed as
\begin{eqnarray}
I_t =\frac{\braket{a^{t}_{\rm out}{}^\dag a^t_{\rm out}}_{\rm ss}}{|z|^2|\alpha|^2F} \!=\! 1 +{\rm Re}\lbrace R G(\omega)\rbrace + {\cal O}\left(|\alpha|^2\right),\label{EqI}
\end{eqnarray}
where the dimensionless complex coefficient $R$ reads
\begin{eqnarray}
R = \frac{\beta}{|z|^2}(\beta-2z).\label{eqR}
\end{eqnarray}
Notice that expression (\ref{EqI}) is valid even in the presence of correlated depashing noise [SM2], and it states that the measurement of $I_t(\omega)$ gives enough information to experimentally obtain the real part of $RG(\omega)$. Since $RG(\omega)$ is an analytic function of $\omega$, we can infer its imaginary part from the Kramers-Kronig relation, ${\rm Im}[R G(\omega)]=\frac{1}{\pi}\mathcal{P}\int d\omega'\frac{ {\rm Re}[R G(\omega')]}{\omega-\omega'}$, and therefore we obtain $G(\omega)$ as
\begin{equation}
G(\omega)=\frac{{\rm Re}[RG(\omega)]+{\rm Im}[RG(\omega)]}{R}
\label{RG}
\end{equation}
Notice that in this method, $\beta$ and $z$ have to be known, and we obtain these parameters from the fit discussed in Sec.~I.E. Finally, using the experimental form of $G(\omega)$ in the expressions (\ref{singlePhotonTransmission}) and (\ref{singlePhotonreflection}), we obtain the reconstructed $t(\omega)$ and $r(\omega)$ that we plot in Fig.~3 of the main text.

\subsection{II.C.- Two-photon reconstruction formulas}

To derive the two-photon reconstruction formulas, we follow the main idea of Ref.~[SM6] and consider an attenuated coherent state in the limit of low excitation power,~i.e. much less than one photon interacting with the QD per lifetime ($|\Omega|^2\ll \gamma_{tot}^2$). In this case, we can expand the coherent state input in a superposition of zero, one, two, and more photons as
\begin{align}
   |\Psi_{\rm in}^{(\alpha)}\rangle = |0\rangle + \alpha |\Psi_{\rm in}^{(1)}\rangle+\frac{\alpha^2}{\sqrt{2}}|\Psi_{\rm in}^{(2)}\rangle+{\cal O}(\alpha^3).\label{weakInputSM}
\end{align}
Here, the single-photon $|\Psi_{\rm in}^{(1)}\rangle$  and two-photon $|\Psi_{\rm in}^{(2)}\rangle$ input wavepackets read
\begin{align}
|\Psi_{\rm in}^{(1)}\rangle ={}& \int d\omega f_{\bar{\omega}}(\omega)a_{\rm in}^t{}^\dag(\omega)|0\rangle,\label{SingleInput}\\
|\Psi_{\rm in}^{(2)}\rangle={}&\iint \frac{d\omega_1 d\omega_2}{\sqrt{2}} f_{\bar{\omega}}(\omega_1)f_{\bar{\omega}}(\omega_2)a_{\rm in}^t{}^\dag(\omega_1)a_{\rm in}^t{}^\dag(\omega_2)|0\rangle,\label{TwoInput}
\end{align}
where $f_{\bar{\omega}}(\omega)$ is the wavepacket shape centered at $\bar{\omega}$, and $|1_\omega^\mu\rangle=a_{\rm in}^\mu{}^\dag(\omega)|0\rangle$ denotes the Fock state of a photon of frequency $\omega$ and propagation direction $\mu=t,r$ along the waveguide.

Following Eq.~(\ref{ScatteringAction}), we apply the scattering matrix $U$ on the input states (\ref{SingleInput}) and (\ref{TwoInput}), and obtain the single-photon $|\Psi_{\rm out}^{(1)}\rangle = U |\Psi_{\rm in}^{(1)}\rangle$ and two-photon $|\Psi_{\rm out}^{(2)}\rangle = U |\Psi_{\rm in}^{(2)}\rangle$ output states, which read
\begin{align}
	|\Psi_{\rm out}^{(1)}\rangle={}&\sum_{\mu=t,r}\iint d\nu d\omega f_{\bar{\omega}}(\omega)S^\mu_{\nu\omega}a_{\rm in}^{\mu}{}^\dag(\nu)|0\rangle,\label{singleOutputExp}\\
	|\Psi_{\rm out}^{(2)}\rangle={}&\sum_{\mu\mu'=t,r}\iint \frac{d\nu_1 d\nu_2}{2\sqrt{2}} \iint d\omega_1 d\omega_2 f_{\bar{\omega}}(\omega_1)f_{\bar{\omega}}(\omega_2)\nonumber\\
	&{}\hspace{1cm}\times S^{\mu\mu'}_{\nu_1\nu_2\omega_1\omega_2}a_{\rm in}^{\mu}{}^\dag(\nu_1)a_{\rm in}^{\mu'}{}^\dag(\nu_2)|0\rangle.\label{twoOutputExp}
\end{align}
By replacing Eqs.~(\ref{s1Def212}) and (\ref{s2Def22}) in Eqs.~(\ref{singleOutputExp}) and (\ref{twoOutputExp}), respectively, we obtain the output states stated in the main text.

On the other hand, if we use the input state (\ref{weakInputSM}) in the definition of the intensity (\ref{Ideff}) and second-order correlation (\ref{g2def}), we can conveniently express them as
\begin{align}
    I_\mu(\omega)={}&|\psi^{(1)}_{\mu}(t)|^2/(|z|^2F)+ {\cal O}\left(|\alpha|^2\right),\\
    g^{(2)}_{\mu\mu'}(t,t+\tau)={}& \frac{\big|\psi^{(2)}_{\mu\mu'}(t,t+\tau)\big|^2}{2\big|\psi^{(1)}_{\mu}(t)\big|^2\big|\psi^{(1)}_{\mu'}(t+\tau)\big|^2} + {\cal O}\left(|\alpha|^2\right),\label{g2psi}
\end{align}
where $\psi^{(1)}_{\mu}(t)= (2\pi)^{-1/2}\int d\nu e^{-i\nu t}\langle 0| a_{\rm in}^\mu(\nu)|\Psi^{(1)}_{\rm out}\rangle$ is the single-photon wavefunction projected on time $t$ and direction $\mu$, and $\psi^{(2)}_{\mu\mu'}(t,t+\tau)= (2\pi)^{-1}\iint d\nu d\nu' e^{-i(\nu t+\nu'[t+\tau])}\langle 0| a_{\rm in}^\mu(\nu)a_{\rm in}^{\mu'}(\nu')|\Psi^{(2)}_{\rm out}\rangle$ is the two-photon wavefunction projected on times $t$ and $t+\tau$, and directions $\mu,\mu'=t,r$.

We can explicitly calculate the single- and two-photon wavefunctions by using Eqs.~(\ref{singleOutputExp})-(\ref{twoOutputExp}) and Eqs.~(\ref{s1Def212})-(\ref{s2Def22}), obtaining
\begin{align}
    \psi^{(1)}_{\mu}(t) = {}&\frac{1}{\sqrt{2\pi}}\int d\omega e^{-i\omega t}f_{\bar{\omega}}(\omega)\chi^\mu(\omega),\label{Psi1}\\
    \psi^{(2)}_{\mu\mu'}(t,t+\tau)={}&\iint \frac{d\omega_1 d\omega_2}{\sqrt{2}\pi } f_{\bar{\omega}}(\omega_1)f_{\bar{\omega}}(\omega_2)e^{-i(\omega_1+\omega_2)(t+\frac{\tau}{2})}\nonumber\\
	\times&{}\left\lbrace e^{\frac{i}{2}(\omega_1-\omega_2)\tau}\chi^{\mu}(\omega_1)\chi^{\mu'}(\omega_2) + {\cal T}_{\omega_1\omega_2}\left(\tau\right)\right\rbrace.\label{Psi2}
\end{align}
Here, we have defined ${\cal T}_{\omega_1\omega_2}(\tau)$ as the Fourier transformed two-photon scattering coefficient and given explicitly as
\begin{equation}
{\cal T}_{\omega_1\omega_2}(\tau)=\frac{1}{2}\int d \Delta e^{-i \Delta\tau}T_{\hat{\omega}-\Delta,\hat{\omega}+\Delta,\omega_1,\omega_2},\label{FTTTTT}
\end{equation}
with $\hat{\omega} = (\omega_1+\omega_2)/2$.

Notice that Eqs.~(\ref{Psi1}) and (\ref{Psi2}) are valid for any wavepacket shape of the input coherent state. In the case of a monochromatic cw input with frequency $\omega$ and constant flux $F$, the wavepacket profile takes the form in Eq.~(\ref{cwlaser}) [SM4], and the single- and two-photon wavefunctions read
\begin{align}
\psi^{(1)}_{\mu}(t)={}&\sqrt{F}e^{-i\omega t}\chi^\mu(\omega),\label{psi1Second}\\ \psi^{(2)}_{\mu\mu'}(t,t+\tau)={}&\sqrt{2}Fe^{-i\omega(2t+\tau)}\lbrace\chi^\mu(\omega)\chi^{\mu'}(\omega)+{\cal T}(\omega,\tau)\rbrace.\label{psi2Second}
\end{align}
with ${\cal T}(\omega,\tau)={\cal T}_{\omega\omega}(\tau)$. Replacing Eqs.~(\ref{psi1Second}) and (\ref{psi2Second}) into Eq.~(\ref{g2psi}), the second-order photo-correlation function reads
\begin{equation}
g^{(2)}_{\mu\mu'} (\tau)=\frac{|\chi^{\mu}(\omega)\chi^{\mu'}(\omega)+\mathcal{T}(\omega,\tau)|^2}{|\chi^{\mu}(\omega)\chi^{\mu'}(\omega)|^2}+ {\cal O}\left(|\alpha|^2\right),\label{g222}
\end{equation}
and this is Eq.~(5) of the main text. Finally, if we expand the square in the numerator of Eq.~(\ref{g222}), one can straightforwardly show that the combination in Eq.~(7) of the main text solves for the correlated part of the scattering : $\mathcal{T}(\omega,\tau)$.

Since the function $T_{\nu_1\nu_2\omega_1\omega_2}$ is analytical for a two-level system, we can use a Kramers-Kronig relation to obtain the imaginary part from the experimentally reconstructed real part. Namely,
\begin{eqnarray}
{\rm Im}[\mathcal{T}(\omega,\tau)]&=\frac{1}{\pi}\mathcal{P}\int d\omega'\frac{Re[\mathcal{T}(\omega',\tau)]}{\omega-\omega'}.
\label{TKrammersSM}
\end{eqnarray}
Therefore, to reconstruct the full $\mathcal{T}(\omega,\tau)$, one needs to acquire $g^{(2)}_{\mu \mu'}$ in the three different configurations, for a set of different input frequencies. If we also sample over different time delays, we can finally obtain the intrinsic two-photon correlations by inverse Fourier transform as $T_{\omega-\Delta,\omega+\Delta,\omega,\omega}=\frac{1}{\pi}\int d\tau e^{i\Delta\tau}\mathcal{T}(\omega,\tau)$.

\subsection{II.D.- Prediction for two-photon correlations in the presence of experimental imperfections}

In Eq.~(3) of the main text we showed the prediction of the two-photon correlations in the case of an ideally coherent TLE with coupling efficiency $\beta$ [SM5]. As we have shown in Sec.~I of this SM, in our experiment there is in addition Fano resonance, white noise dephasing, spectral diffusion, and instrument response function, which will modify this prediction. Therefore, in this subsection we derive the theoretical prediction for the two-photon correlation functions in the presence of the imperfections, and which we use in Fig.~4(b) to verify our experimental reconstruction.

First, we calculate the two-photon scattering matrix (\ref{s2Def1}) using the standard input-ouput theory method [SM5] and the quantum Langevin equations (\ref{EqMotSm})-(\ref{EqMotSz}), which incorporate the effects of Fano resonance $z$ and white noise dephasing $\gamma_d$. We find,
\begin{align}
    T_{\nu_1\nu_2\omega_1\omega_2} = -\frac{4\beta^2}{\pi\gamma_{tot}}\frac{z^4}{|z|^4}\frac{G(\nu_1)G(\nu_2)G(\omega_1)G(\omega_2)}{G(\frac{\omega_1+\omega_2}{2})},
\end{align}
where the function $G(\omega)$ is the one appearing in the single-photon scattering expression in Eq.~(\ref{EqG}).

Since our method uses a monochromatic input laser with frequency $\omega$, and the TLE conserves the total energy, our method is only sensitive to reconstruct the $T_{\omega-\Delta,\omega+\Delta,\omega,\omega}$ sector of the two-photon nonlinearity. Using the explicit form of $G(\omega)$ in Eq.~(\ref{EqG}), we find
\begin{equation}
T_{\omega-\Delta,\omega+\Delta,\omega,\omega}=-\frac{4\beta^2}{\pi \gamma_{tot}}\frac{z^4}{|z|^4}\left[ \frac{G(\omega)^3}{1+(2G(\omega)\Delta/\gamma_{tot})^2} \right].\label{nonlinear2}
\end{equation}

From $T_{\omega-\Delta,\omega+\Delta,\omega,\omega}$ is Eq.~(\ref{nonlinear2}), we can calculate its Fourier transform using Eq.~(\ref{FTTTTT}), obtaining
\begin{equation}
\mathcal{T}(\omega,\tau)=-\beta^2\frac{z^4}{|z|^4}G(\omega)^2 e^{-(\gamma_{tot}/2+\gamma_{d}-i[\omega-\omega_0])|\tau|}\label{IdealTTT}
\end{equation}

Finally, to account for spectral diffusion and IRF, we average the above prediction over the same distributions discussed in Sec.~I.D, obtaining
\begin{equation}
\bar{\mathcal{T}}(\omega,\tau)=\int d\Delta \int d\tau' P_{\mathrm{SD}}(\Delta,\sigma_{\mathrm{long}})P_{\mathrm{IRF}}(\tau'-\tau)\mathcal{T}(\omega-\Delta,\tau'),
\end{equation}
with $P_{\mathrm{SD}}(\Delta,\sigma_{\mathrm{long}})$ and $P_{IRF}(\tau')$ given in Eqs.~(\ref{GaussSD}) and (\ref{GaussIRF}), respectively. The real part of this expression $\bar{\mathcal{T}}(\omega,\tau)$, wich includes all experimental imperfections, is the quantity that we verify in Fig.~4(b) when comparing it to the experimentally reconstructed two-photon correlations.

\section*{References}

\begin{enumerate}[label={[SM\arabic*]}]
    \item C.~W. Gardiner, and P. Zoller, Quantum Noise (Springer Verlag, Berlin, 3rd Ed., 2004).
    \item T. Ramos, and J.~J. Garc\'ia-Ripoll, New J. Phys. 20, 105007 (2018).
    \item A. Auff\`eves-Garnier, C. Simon, J.-~M. G\'erard, and J.-~P. Poizat, Phys. Rev. A 75, 053823 (2007).
    \item K. J. Blow, R. Loudon, S.~J.~D. Phoenix, and T.~J. Shepherd, Phys Rev. A 42, 4102 (1990).
    \item S. Fan, S.~E. Kocaba\c{s}, and J.~-T. Shen, Phys. Rev. A 82, 063821 (2010).
    \item T. Ramos, and J.~J. Garc\'ia-Ripoll, Phys. Rev. Lett. 119, 153601 (2017).
\end{enumerate}
\

\begin{thebibliography}{55}%
\makeatletter
\providecommand \@ifxundefined [1]{%
 \@ifx{#1\undefined}
}%
\providecommand \@ifnum [1]{%
 \ifnum #1\expandafter \@firstoftwo
 \else \expandafter \@secondoftwo
 \fi
}%
\providecommand \@ifx [1]{%
 \ifx #1\expandafter \@firstoftwo
 \else \expandafter \@secondoftwo
 \fi
}%
\providecommand \natexlab [1]{#1}%
\providecommand \enquote  [1]{``#1''}%
\providecommand \bibnamefont  [1]{#1}%
\providecommand \bibfnamefont [1]{#1}%
\providecommand \citenamefont [1]{#1}%
\providecommand \href@noop [0]{\@secondoftwo}%
\providecommand \href [0]{\begingroup \@sanitize@url \@href}%
\providecommand \@href[1]{\@@startlink{#1}\@@href}%
\providecommand \@@href[1]{\endgroup#1\@@endlink}%
\providecommand \@sanitize@url [0]{\catcode `\\12\catcode `\$12\catcode
  `\&12\catcode `\#12\catcode `\^12\catcode `\_12\catcode `\%12\relax}%
\providecommand \@@startlink[1]{}%
\providecommand \@@endlink[0]{}%
\providecommand \url  [0]{\begingroup\@sanitize@url \@url }%
\providecommand \@url [1]{\endgroup\@href {#1}{\urlprefix }}%
\providecommand \urlprefix  [0]{URL }%
\providecommand \Eprint [0]{\href }%
\providecommand \doibase [0]{http://dx.doi.org/}%
\providecommand \selectlanguage [0]{\@gobble}%
\providecommand \bibinfo  [0]{\@secondoftwo}%
\providecommand \bibfield  [0]{\@secondoftwo}%
\providecommand \translation [1]{[#1]}%
\providecommand \BibitemOpen [0]{}%
\providecommand \bibitemStop [0]{}%
\providecommand \bibitemNoStop [0]{.\EOS\space}%
\providecommand \EOS [0]{\spacefactor3000\relax}%
\providecommand \BibitemShut  [1]{\csname bibitem#1\endcsname}%
\let\auto@bib@innerbib\@empty
%</preamble>
\bibitem [{\citenamefont {Ladd}\ \emph {et~al.}(2010)\citenamefont {Ladd},
  \citenamefont {Jelezko}, \citenamefont {Laflamme}, \citenamefont {Nakamura},
  \citenamefont {Monroe},\ and\ \citenamefont {O'Brien}}]{Ladd_2010}%
  \BibitemOpen
  \bibfield  {author} {\bibinfo {author} {\bibfnamefont {T.~D.}\ \bibnamefont
  {Ladd}}, \bibinfo {author} {\bibfnamefont {F.}~\bibnamefont {Jelezko}},
  \bibinfo {author} {\bibfnamefont {R.}~\bibnamefont {Laflamme}}, \bibinfo
  {author} {\bibfnamefont {Y.}~\bibnamefont {Nakamura}}, \bibinfo {author}
  {\bibfnamefont {C.}~\bibnamefont {Monroe}},\ and\ \bibinfo {author}
  {\bibfnamefont {J.~L.}\ \bibnamefont {O'Brien}},\ }\href {\doibase
  10.1038/nature08812} {\bibfield  {journal} {\bibinfo  {journal} {Nature}\
  }\textbf {\bibinfo {volume} {464}},\ \bibinfo {pages} {45–53} (\bibinfo
  {year} {2010})}\BibitemShut {NoStop}%
\bibitem [{\citenamefont {Chang}\ \emph {et~al.}(2014)\citenamefont {Chang},
  \citenamefont {Vuleti{\'{c}}},\ and\ \citenamefont {Lukin}}]{Chang_2014}%
  \BibitemOpen
  \bibfield  {author} {\bibinfo {author} {\bibfnamefont {D.~E.}\ \bibnamefont
  {Chang}}, \bibinfo {author} {\bibfnamefont {V.}~\bibnamefont
  {Vuleti{\'{c}}}},\ and\ \bibinfo {author} {\bibfnamefont {M.~D.}\
  \bibnamefont {Lukin}},\ }\href {\doibase 10.1038/nphoton.2014.192} {\bibfield
   {journal} {\bibinfo  {journal} {Nature Photonics}\ }\textbf {\bibinfo
  {volume} {8}},\ \bibinfo {pages} {685} (\bibinfo {year} {2014})}\BibitemShut
  {NoStop}%
\bibitem [{\citenamefont {Lodahl}\ \emph {et~al.}(2015)\citenamefont {Lodahl},
  \citenamefont {Mahmoodian},\ and\ \citenamefont {Stobbe}}]{Lodahl_2015}%
  \BibitemOpen
  \bibfield  {author} {\bibinfo {author} {\bibfnamefont {P.}~\bibnamefont
  {Lodahl}}, \bibinfo {author} {\bibfnamefont {S.}~\bibnamefont {Mahmoodian}},
 \ and\ \bibinfo {author} {\bibfnamefont {S.}~\bibnamefont {Stobbe}},\ }\href
  {\doibase 10.1103/RevModPhys.87.347} {\bibfield  {journal} {\bibinfo
  {journal} {Rev. Mod. Phys.}\ }\textbf {\bibinfo {volume} {87}},\ \bibinfo
  {pages} {347} (\bibinfo {year} {2015})}\BibitemShut {NoStop}%
\bibitem [{\citenamefont {Tiarks}\ \emph {et~al.}(2018)\citenamefont {Tiarks},
  \citenamefont {Schmidt-Eberle}, \citenamefont {Stolz}, \citenamefont
  {Rempe},\ and\ \citenamefont {D\"{u}rr}}]{Tiarks_2018}%
  \BibitemOpen
  \bibfield  {author} {\bibinfo {author} {\bibfnamefont {D.}~\bibnamefont
  {Tiarks}}, \bibinfo {author} {\bibfnamefont {S.}~\bibnamefont
  {Schmidt-Eberle}}, \bibinfo {author} {\bibfnamefont {T.}~\bibnamefont
  {Stolz}}, \bibinfo {author} {\bibfnamefont {G.}~\bibnamefont {Rempe}},\ and\
  \bibinfo {author} {\bibfnamefont {S.}~\bibnamefont {D\"{u}rr}},\ }\href
  {\doibase 10.1038/s41567-018-0313-7} {\bibfield  {journal} {\bibinfo
  {journal} {Nature Physics}\ }\textbf {\bibinfo {volume} {15}},\ \bibinfo
  {pages} {124-–126} (\bibinfo {year} {2018})}\BibitemShut {NoStop}%
\bibitem [{\citenamefont {Baur}\ \emph {et~al.}(2014)\citenamefont {Baur},
  \citenamefont {Tiarks}, \citenamefont {Rempe},\ and\ \citenamefont
  {D\"{u}rr}}]{Baur_2014}%
  \BibitemOpen
  \bibfield  {author} {\bibinfo {author} {\bibfnamefont {S.}~\bibnamefont
  {Baur}}, \bibinfo {author} {\bibfnamefont {D.}~\bibnamefont {Tiarks}},
  \bibinfo {author} {\bibfnamefont {G.}~\bibnamefont {Rempe}},\ and\ \bibinfo
  {author} {\bibfnamefont {S.}~\bibnamefont {D\"{u}rr}},\ }\href {\doibase
  10.1103/PhysRevLett.112.073901} {\bibfield  {journal} {\bibinfo  {journal}
  {Phys. Rev. Lett.}\ }\textbf {\bibinfo {volume} {112}},\ \bibinfo {pages}
  {073901} (\bibinfo {year} {2014})}\BibitemShut {NoStop}%
\bibitem [{\citenamefont {Reiserer}\ \emph {et~al.}(2014)\citenamefont
  {Reiserer}, \citenamefont {Kalb}, \citenamefont {Rempe},\ and\ \citenamefont
  {Ritter}}]{Reiserer_2014}%
  \BibitemOpen
  \bibfield  {author} {\bibinfo {author} {\bibfnamefont {A.}~\bibnamefont
  {Reiserer}}, \bibinfo {author} {\bibfnamefont {N.}~\bibnamefont {Kalb}},
  \bibinfo {author} {\bibfnamefont {G.}~\bibnamefont {Rempe}},\ and\ \bibinfo
  {author} {\bibfnamefont {S.}~\bibnamefont {Ritter}},\ }\href {\doibase
  10.1038/nature13177} {\bibfield  {journal} {\bibinfo  {journal} {Nature}\
  }\textbf {\bibinfo {volume} {508}},\ \bibinfo {pages} {237–240} (\bibinfo
  {year} {2014})}\BibitemShut {NoStop}%
\bibitem [{\citenamefont {Najer}\ \emph {et~al.}(2019)\citenamefont {Najer},
  \citenamefont {S\"{o}llner}, \citenamefont {Sekatski}, \citenamefont
  {Dolique}, \citenamefont {L\"{o}bl}, \citenamefont {Riedel}, \citenamefont
  {Schott}, \citenamefont {Starosielec}, \citenamefont {Valentin},
  \citenamefont {Wieck}, \citenamefont {Sangouard}, \citenamefont {Ludwig},\
  and\ \citenamefont {Warburton}}]{Najer_2019}%
  \BibitemOpen
  \bibfield  {author} {\bibinfo {author} {\bibfnamefont {D.}~\bibnamefont
  {Najer}}, \bibinfo {author} {\bibfnamefont {I.}~\bibnamefont {S\"{o}llner}},
  \bibinfo {author} {\bibfnamefont {P.}~\bibnamefont {Sekatski}}, \bibinfo
  {author} {\bibfnamefont {V.}~\bibnamefont {Dolique}}, \bibinfo {author}
  {\bibfnamefont {M.~C.}\ \bibnamefont {L\"{o}bl}}, \bibinfo {author}
  {\bibfnamefont {D.}~\bibnamefont {Riedel}}, \bibinfo {author} {\bibfnamefont
  {R.}~\bibnamefont {Schott}}, \bibinfo {author} {\bibfnamefont
  {S.}~\bibnamefont {Starosielec}}, \bibinfo {author} {\bibfnamefont {S.~R.}\
  \bibnamefont {Valentin}}, \bibinfo {author} {\bibfnamefont {A.~D.}\
  \bibnamefont {Wieck}}, \bibinfo {author} {\bibfnamefont {N.}~\bibnamefont
  {Sangouard}}, \bibinfo {author} {\bibfnamefont {A.}~\bibnamefont {Ludwig}}, \
  and\ \bibinfo {author} {\bibfnamefont {R.~J.}\ \bibnamefont {Warburton}},\
  }\href {\doibase 10.1038/s41586-019-1709-y} {\bibfield  {journal} {\bibinfo
  {journal} {Nature}\ }\textbf {\bibinfo {volume} {575}},\ \bibinfo {pages}
  {622} (\bibinfo {year} {2019})}\BibitemShut {NoStop}%
\bibitem [{\citenamefont {Deppe}\ \emph {et~al.}(2008)\citenamefont {Deppe},
  \citenamefont {Mariantoni}, \citenamefont {Menzel}, \citenamefont {Marx},
  \citenamefont {Saito}, \citenamefont {Kakuyanagi}, \citenamefont {Tanaka},
  \citenamefont {Meno}, \citenamefont {Semba}, \citenamefont {Takayanagi},\
  and\ \citenamefont {et~al.}}]{Deppe_2008}%
  \BibitemOpen
  \bibfield  {author} {\bibinfo {author} {\bibfnamefont {F.}~\bibnamefont
  {Deppe}}, \bibinfo {author} {\bibfnamefont {M.}~\bibnamefont {Mariantoni}},
  \bibinfo {author} {\bibfnamefont {E.~P.}\ \bibnamefont {Menzel}}, \bibinfo
  {author} {\bibfnamefont {A.}~\bibnamefont {Marx}}, \bibinfo {author}
  {\bibfnamefont {S.}~\bibnamefont {Saito}}, \bibinfo {author} {\bibfnamefont
  {K.}~\bibnamefont {Kakuyanagi}}, \bibinfo {author} {\bibfnamefont
  {H.}~\bibnamefont {Tanaka}}, \bibinfo {author} {\bibfnamefont
  {T.}~\bibnamefont {Meno}}, \bibinfo {author} {\bibfnamefont {K.}~\bibnamefont
  {Semba}}, \bibinfo {author} {\bibfnamefont {H.}~\bibnamefont {Takayanagi}}, \
  and\ \bibinfo {author} {\bibnamefont {et~al.}},\ }\href {\doibase
  10.1038/nphys1016} {\bibfield  {journal} {\bibinfo  {journal} {Nature
  Physics}\ }\textbf {\bibinfo {volume} {4}},\ \bibinfo {pages} {686–691}
  (\bibinfo {year} {2008})}\BibitemShut {NoStop}%
\bibitem [{\citenamefont {Mirhosseini}\ \emph {et~al.}(2019)\citenamefont
  {Mirhosseini}, \citenamefont {Kim}, \citenamefont {Zhang}, \citenamefont
  {Sipahigil}, \citenamefont {Dieterle}, \citenamefont {Keeler}, \citenamefont
  {Asenjo-Garcia}, \citenamefont {Chang},\ and\ \citenamefont
  {Painter}}]{Mirhosseini_2019}%
  \BibitemOpen
  \bibfield  {author} {\bibinfo {author} {\bibfnamefont {M.}~\bibnamefont
  {Mirhosseini}}, \bibinfo {author} {\bibfnamefont {E.}~\bibnamefont {Kim}},
  \bibinfo {author} {\bibfnamefont {X.}~\bibnamefont {Zhang}}, \bibinfo
  {author} {\bibfnamefont {A.}~\bibnamefont {Sipahigil}}, \bibinfo {author}
  {\bibfnamefont {P.~B.}\ \bibnamefont {Dieterle}}, \bibinfo {author}
  {\bibfnamefont {A.~J.}\ \bibnamefont {Keeler}}, \bibinfo {author}
  {\bibfnamefont {A.}~\bibnamefont {Asenjo-Garcia}}, \bibinfo {author}
  {\bibfnamefont {D.~E.}\ \bibnamefont {Chang}},\ and\ \bibinfo {author}
  {\bibfnamefont {O.}~\bibnamefont {Painter}},\ }\href {\doibase
  10.1038/s41586-019-1196-1} {\bibfield  {journal} {\bibinfo  {journal}
  {Nature}\ }\textbf {\bibinfo {volume} {569}},\ \bibinfo {pages} {692–}
  (\bibinfo {year} {2019})}\BibitemShut {NoStop}%
\bibitem [{\citenamefont {Lang}\ \emph {et~al.}(2011)\citenamefont {Lang},
  \citenamefont {Bozyigit}, \citenamefont {Eichler}, \citenamefont {Steffen},
  \citenamefont {Fink}, \citenamefont {Abdumalikov}, \citenamefont {Baur},
  \citenamefont {Filipp}, \citenamefont {da~Silva}, \citenamefont {Blais},\
  and\ \citenamefont {Wallraff}}]{Lang_2011}%
  \BibitemOpen
  \bibfield  {author} {\bibinfo {author} {\bibfnamefont {C.}~\bibnamefont
  {Lang}}, \bibinfo {author} {\bibfnamefont {D.}~\bibnamefont {Bozyigit}},
  \bibinfo {author} {\bibfnamefont {C.}~\bibnamefont {Eichler}}, \bibinfo
  {author} {\bibfnamefont {L.}~\bibnamefont {Steffen}}, \bibinfo {author}
  {\bibfnamefont {J.~M.}\ \bibnamefont {Fink}}, \bibinfo {author}
  {\bibfnamefont {A.~A.}\ \bibnamefont {Abdumalikov}}, \bibinfo {author}
  {\bibfnamefont {M.}~\bibnamefont {Baur}}, \bibinfo {author} {\bibfnamefont
  {S.}~\bibnamefont {Filipp}}, \bibinfo {author} {\bibfnamefont {M.~P.}\
  \bibnamefont {da~Silva}}, \bibinfo {author} {\bibfnamefont {A.}~\bibnamefont
  {Blais}},\ and\ \bibinfo {author} {\bibfnamefont {A.}~\bibnamefont
  {Wallraff}},\ }\href {\doibase 10.1103/PhysRevLett.106.243601} {\bibfield
  {journal} {\bibinfo  {journal} {Phys. Rev. Lett.}\ }\textbf {\bibinfo
  {volume} {106}},\ \bibinfo {pages} {243601} (\bibinfo {year}
  {2011})}\BibitemShut {NoStop}%
\bibitem [{\citenamefont {Bajcsy}\ \emph {et~al.}(2009)\citenamefont {Bajcsy},
  \citenamefont {Hofferberth}, \citenamefont {Balic}, \citenamefont {Peyronel},
  \citenamefont {Hafezi}, \citenamefont {Zibrov}, \citenamefont {Vuleti\'{c}},\
  and\ \citenamefont {Lukin}}]{Bajcsy_2009}%
  \BibitemOpen
  \bibfield  {author} {\bibinfo {author} {\bibfnamefont {M.}~\bibnamefont
  {Bajcsy}}, \bibinfo {author} {\bibfnamefont {S.}~\bibnamefont {Hofferberth}},
  \bibinfo {author} {\bibfnamefont {V.}~\bibnamefont {Balic}}, \bibinfo
  {author} {\bibfnamefont {T.}~\bibnamefont {Peyronel}}, \bibinfo {author}
  {\bibfnamefont {M.}~\bibnamefont {Hafezi}}, \bibinfo {author} {\bibfnamefont
  {A.~S.}\ \bibnamefont {Zibrov}}, \bibinfo {author} {\bibfnamefont
  {V.}~\bibnamefont {Vuleti\'{c}}},\ and\ \bibinfo {author} {\bibfnamefont
  {M.~D.}\ \bibnamefont {Lukin}},\ }\href {\doibase
  10.1103/PhysRevLett.102.203902} {\bibfield  {journal} {\bibinfo  {journal}
  {Phys. Rev. Lett.}\ }\textbf {\bibinfo {volume} {102}},\ \bibinfo {pages}
  {203902} (\bibinfo {year} {2009})}\BibitemShut {NoStop}%
\bibitem [{\citenamefont {Prasad}\ \emph {et~al.}(2019)\citenamefont {Prasad},
  \citenamefont {Hinney}, \citenamefont {Mahmoodian}, \citenamefont {Hammerer},
  \citenamefont {Rind}, \citenamefont {Schneeweiss}, \citenamefont
  {S\o{}rensen}, \citenamefont {Volz},\ and\ \citenamefont
  {Rauschenbeutel}}]{Prasad_2019}%
  \BibitemOpen
  \bibfield  {author} {\bibinfo {author} {\bibfnamefont {A.~S.}\ \bibnamefont
  {Prasad}}, \bibinfo {author} {\bibfnamefont {J.}~\bibnamefont {Hinney}},
  \bibinfo {author} {\bibfnamefont {S.}~\bibnamefont {Mahmoodian}}, \bibinfo
  {author} {\bibfnamefont {K.}~\bibnamefont {Hammerer}}, \bibinfo {author}
  {\bibfnamefont {S.}~\bibnamefont {Rind}}, \bibinfo {author} {\bibfnamefont
  {P.}~\bibnamefont {Schneeweiss}}, \bibinfo {author} {\bibfnamefont {A.~S.}\
  \bibnamefont {S\o{}rensen}}, \bibinfo {author} {\bibfnamefont
  {J.}~\bibnamefont {Volz}},\ and\ \bibinfo {author} {\bibfnamefont
  {A.}~\bibnamefont {Rauschenbeutel}},\ }\href@noop {} {} (\bibinfo {year}
  {2019}),\ \Eprint {http://arxiv.org/abs/1911.09701} {arXiv:1911.09701
  [quant-ph]} \BibitemShut {NoStop}%
\bibitem [{\citenamefont {Goban}\ \emph {et~al.}(2014)\citenamefont {Goban},
  \citenamefont {Hung}, \citenamefont {Yu}, \citenamefont {Hood}, \citenamefont
  {Muniz}, \citenamefont {Lee}, \citenamefont {Martin}, \citenamefont
  {McClung}, \citenamefont {Choi}, \citenamefont {Chang}, \citenamefont
  {Painter},\ and\ \citenamefont {Kimble}}]{Goban_2014}%
  \BibitemOpen
  \bibfield  {author} {\bibinfo {author} {\bibfnamefont {A.}~\bibnamefont
  {Goban}}, \bibinfo {author} {\bibfnamefont {C.-L.}\ \bibnamefont {Hung}},
  \bibinfo {author} {\bibfnamefont {S.-P.}\ \bibnamefont {Yu}}, \bibinfo
  {author} {\bibfnamefont {J.}~\bibnamefont {Hood}}, \bibinfo {author}
  {\bibfnamefont {J.}~\bibnamefont {Muniz}}, \bibinfo {author} {\bibfnamefont
  {J.}~\bibnamefont {Lee}}, \bibinfo {author} {\bibfnamefont {M.}~\bibnamefont
  {Martin}}, \bibinfo {author} {\bibfnamefont {A.}~\bibnamefont {McClung}},
  \bibinfo {author} {\bibfnamefont {K.}~\bibnamefont {Choi}}, \bibinfo {author}
  {\bibfnamefont {D.}~\bibnamefont {Chang}}, \bibinfo {author} {\bibfnamefont
  {O.}~\bibnamefont {Painter}},\ and\ \bibinfo {author} {\bibfnamefont
  {H.~J.}\ \bibnamefont {Kimble}},\ }\href
  {http://dx.doi.org/10.1038/ncomms4808} {\bibfield  {journal} {\bibinfo
  {journal} {Nat. Commun.}\ }\textbf {\bibinfo {volume} {5}} ,\ \bibinfo
  {pages} {3808} (\bibinfo
  {year} {2014})}\BibitemShut {NoStop}%
\bibitem [{\citenamefont {Javadi}\ \emph {et~al.}(2015)\citenamefont {Javadi},
  \citenamefont {S\"{o}llner}, \citenamefont {Arcari}, \citenamefont {Hansen},
  \citenamefont {Midolo}, \citenamefont {Mahmoodian}, \citenamefont
  {Kir{\v{s}}ansk{\.{e}}}, \citenamefont {Pregnolato}, \citenamefont {Lee},
  \citenamefont {Song},\ and\ \citenamefont {et~al.}}]{Javadi_2015}%
  \BibitemOpen
  \bibfield  {author} {\bibinfo {author} {\bibfnamefont {A.}~\bibnamefont
  {Javadi}}, \bibinfo {author} {\bibfnamefont {I.}~\bibnamefont {S\"{o}llner}},
  \bibinfo {author} {\bibfnamefont {M.}~\bibnamefont {Arcari}}, \bibinfo
  {author} {\bibfnamefont {S.~L.}\ \bibnamefont {Hansen}}, \bibinfo {author}
  {\bibfnamefont {L.}~\bibnamefont {Midolo}}, \bibinfo {author} {\bibfnamefont
  {S.}~\bibnamefont {Mahmoodian}}, \bibinfo {author} {\bibfnamefont
  {G.}~\bibnamefont {Kir{\v{s}}ansk{\.{e}}}}, \bibinfo {author} {\bibfnamefont
  {T.}~\bibnamefont {Pregnolato}}, \bibinfo {author} {\bibfnamefont {E.~H.}\
  \bibnamefont {Lee}}, \bibinfo {author} {\bibfnamefont {J.~D.}\ \bibnamefont
  {Song}},\ and\ \bibinfo {author} {\bibnamefont {et~al.}},\ }\href
  {http://dx.doi.org/10.1038/ncomms9655} {\bibfield  {journal} {\bibinfo
  {journal} {Nat. Commun.}\ }\textbf {\bibinfo {volume} {6}},\ \bibinfo
  {pages} {8655} (\bibinfo
  {year} {2015})}\BibitemShut {NoStop}%
\bibitem [{\citenamefont {Thyrrestrup}\ \emph {et~al.}(2018)\citenamefont
  {Thyrrestrup}, \citenamefont {Kir{\v{s}}ansk{\.{e}}}, \citenamefont
  {Le~Jeannic}, \citenamefont {Pregnolato}, \citenamefont {Zhai}, \citenamefont
  {Raahauge}, \citenamefont {Midolo}, \citenamefont {Rotenberg}, \citenamefont
  {Javadi}, \citenamefont {Schott}, \citenamefont {Wieck}, \citenamefont
  {Ludwig}, \citenamefont {L\"{o}bl}, \citenamefont {S\"{o}llner},
  \citenamefont {Warburton},\ and\ \citenamefont {Lodahl}}]{Nielsen_2018}%
  \BibitemOpen
  \bibfield  {author} {\bibinfo {author} {\bibfnamefont {H.}~\bibnamefont
  {Thyrrestrup}}, \bibinfo {author} {\bibfnamefont {G.}~\bibnamefont
  {Kir{\v{s}}ansk{\.{e}}}}, \bibinfo {author} {\bibfnamefont {H.}~\bibnamefont
  {Le~Jeannic}}, \bibinfo {author} {\bibfnamefont {T.}~\bibnamefont
  {Pregnolato}}, \bibinfo {author} {\bibfnamefont {L.}~\bibnamefont {Zhai}},
  \bibinfo {author} {\bibfnamefont {L.}~\bibnamefont {Raahauge}}, \bibinfo
  {author} {\bibfnamefont {L.}~\bibnamefont {Midolo}}, \bibinfo {author}
  {\bibfnamefont {N.}~\bibnamefont {Rotenberg}}, \bibinfo {author}
  {\bibfnamefont {A.}~\bibnamefont {Javadi}}, \bibinfo {author} {\bibfnamefont
  {R.}~\bibnamefont {Schott}}, \bibinfo {author} {\bibfnamefont {A.~D.}\
  \bibnamefont {Wieck}}, \bibinfo {author} {\bibfnamefont {A.}~\bibnamefont
  {Ludwig}}, \bibinfo {author} {\bibfnamefont {M.~C.}\ \bibnamefont
  {L\"{o}bl}}, \bibinfo {author} {\bibfnamefont {I.}~\bibnamefont
  {S\"{o}llner}}, \bibinfo {author} {\bibfnamefont {R.~J.}\ \bibnamefont
  {Warburton}},\ and\ \bibinfo {author} {\bibfnamefont {P.}~\bibnamefont
  {Lodahl}},\ }\href {\doibase 10.1021/acs.nanolett.7b05016} {\bibfield
  {journal} {\bibinfo  {journal} {Nano Letters}\ }\textbf {\bibinfo {volume}
  {18}},\ \bibinfo {pages} {1801} (\bibinfo {year} {2018})}\BibitemShut
  {NoStop}%
\bibitem [{\citenamefont {Hallett}\ \emph {et~al.}(2018)\citenamefont
  {Hallett}, \citenamefont {Foster}, \citenamefont {Hurst}, \citenamefont
  {Royall}, \citenamefont {Kok}, \citenamefont {Clarke}, \citenamefont
  {Itskevich}, \citenamefont {Fox}, \citenamefont {Skolnick},\ and\
  \citenamefont {Wilson}}]{Hallett_2018}%
  \BibitemOpen
  \bibfield  {author} {\bibinfo {author} {\bibfnamefont {D.}~\bibnamefont
  {Hallett}}, \bibinfo {author} {\bibfnamefont {A.~P.}\ \bibnamefont {Foster}},
  \bibinfo {author} {\bibfnamefont {D.~L.}\ \bibnamefont {Hurst}}, \bibinfo
  {author} {\bibfnamefont {B.}~\bibnamefont {Royall}}, \bibinfo {author}
  {\bibfnamefont {P.}~\bibnamefont {Kok}}, \bibinfo {author} {\bibfnamefont
  {E.}~\bibnamefont {Clarke}}, \bibinfo {author} {\bibfnamefont {I.~E.}\
  \bibnamefont {Itskevich}}, \bibinfo {author} {\bibfnamefont {A.~M.}\
  \bibnamefont {Fox}}, \bibinfo {author} {\bibfnamefont {M.~S.}\ \bibnamefont
  {Skolnick}},\ and\ \bibinfo {author} {\bibfnamefont {L.~R.}\ \bibnamefont
  {Wilson}},\ }\href {\doibase 10.1364/OPTICA.5.000644} {\bibfield  {journal}
  {\bibinfo  {journal} {Optica}\ }\textbf {\bibinfo {volume} {5}},\ \bibinfo
  {pages} {644} (\bibinfo {year} {2018})}\BibitemShut {NoStop}%
\bibitem [{\citenamefont {Foster}\ \emph {et~al.}(2019)\citenamefont {Foster},
  \citenamefont {Hallett}, \citenamefont {Iorsh}, \citenamefont {Sheldon},
  \citenamefont {Godsland}, \citenamefont {Royall}, \citenamefont {Clarke},
  \citenamefont {Shelykh}, \citenamefont {Fox}, \citenamefont {Skolnick},
  \citenamefont {Itskevich},\ and\ \citenamefont {Wilson}}]{Foster2019}%
  \BibitemOpen
  \bibfield  {author} {\bibinfo {author} {\bibfnamefont {A.~P.}\ \bibnamefont
  {Foster}}, \bibinfo {author} {\bibfnamefont {D.}~\bibnamefont {Hallett}},
  \bibinfo {author} {\bibfnamefont {I.~V.}\ \bibnamefont {Iorsh}}, \bibinfo
  {author} {\bibfnamefont {S.~J.}\ \bibnamefont {Sheldon}}, \bibinfo {author}
  {\bibfnamefont {M.~R.}\ \bibnamefont {Godsland}}, \bibinfo {author}
  {\bibfnamefont {B.}~\bibnamefont {Royall}}, \bibinfo {author} {\bibfnamefont
  {E.}~\bibnamefont {Clarke}}, \bibinfo {author} {\bibfnamefont {I.~A.}\
  \bibnamefont {Shelykh}}, \bibinfo {author} {\bibfnamefont {A.~M.}\
  \bibnamefont {Fox}}, \bibinfo {author} {\bibfnamefont {M.~S.}\ \bibnamefont
  {Skolnick}}, \bibinfo {author} {\bibfnamefont {I.~E.}\ \bibnamefont
  {Itskevich}},\ and\ \bibinfo {author} {\bibfnamefont {L.~R.}\ \bibnamefont
  {Wilson}},\ }\href {\doibase 10.1103/PhysRevLett.122.173603} {\bibfield
  {journal} {\bibinfo  {journal} {Phys. Rev. Lett.}\ }\textbf {\bibinfo
  {volume} {122}},\ \bibinfo {pages} {173603} (\bibinfo {year}
  {2019})}\BibitemShut {NoStop}%
\bibitem [{\citenamefont {Sipahigil}\ \emph {et~al.}(2016)\citenamefont
  {Sipahigil}, \citenamefont {Evans}, \citenamefont {Sukachev}, \citenamefont
  {Burek}, \citenamefont {Borregaard}, \citenamefont {Bhaskar}, \citenamefont
  {Nguyen}, \citenamefont {Pacheco}, \citenamefont {Atikian}, \citenamefont
  {Meuwly}, \citenamefont {Camacho}, \citenamefont {Jelezko}, \citenamefont
  {Bielejec}, \citenamefont {Park}, \citenamefont {Lon{\v c}ar},\ and\
  \citenamefont {Lukin}}]{Sipahigil_2016}%
  \BibitemOpen
  \bibfield  {author} {\bibinfo {author} {\bibfnamefont {A.}~\bibnamefont
  {Sipahigil}}, \bibinfo {author} {\bibfnamefont {R.~E.}\ \bibnamefont
  {Evans}}, \bibinfo {author} {\bibfnamefont {D.~D.}\ \bibnamefont {Sukachev}},
  \bibinfo {author} {\bibfnamefont {M.~J.}\ \bibnamefont {Burek}}, \bibinfo
  {author} {\bibfnamefont {J.}~\bibnamefont {Borregaard}}, \bibinfo {author}
  {\bibfnamefont {M.~K.}\ \bibnamefont {Bhaskar}}, \bibinfo {author}
  {\bibfnamefont {C.~T.}\ \bibnamefont {Nguyen}}, \bibinfo {author}
  {\bibfnamefont {J.~L.}\ \bibnamefont {Pacheco}}, \bibinfo {author}
  {\bibfnamefont {H.~A.}\ \bibnamefont {Atikian}}, \bibinfo {author}
  {\bibfnamefont {C.}~\bibnamefont {Meuwly}}, \bibinfo {author} {\bibfnamefont
  {R.~M.}\ \bibnamefont {Camacho}}, \bibinfo {author} {\bibfnamefont
  {F.}~\bibnamefont {Jelezko}}, \bibinfo {author} {\bibfnamefont
  {E.}~\bibnamefont {Bielejec}}, \bibinfo {author} {\bibfnamefont
  {H.}~\bibnamefont {Park}}, \bibinfo {author} {\bibfnamefont {M.}~\bibnamefont
  {Lon{\v c}ar}},\ and\ \bibinfo {author} {\bibfnamefont {M.~D.}\ \bibnamefont
  {Lukin}},\ }\href {\doibase 10.1126/science.aah6875} {\bibfield  {journal}
  {\bibinfo  {journal} {Science}\ }\textbf {\bibinfo {volume} {354}},\ \bibinfo
  {pages} {847} (\bibinfo {year} {2016})}\BibitemShut {NoStop}%
\bibitem [{\citenamefont {Uppu}\ \emph {et~al.}(2020)\citenamefont {Uppu},
  \citenamefont {Pedersen}, \citenamefont {Wang}, \citenamefont {Olesen},
  \citenamefont {Papon}, \citenamefont {Zhou}, \citenamefont {Midolo},
  \citenamefont {Scholz}, \citenamefont {Wieck}, \citenamefont {Ludwig},\ and\
  \citenamefont {Lodahl}}]{Uppu_2020}%
  \BibitemOpen
  \bibfield  {author} {\bibinfo {author} {\bibfnamefont {R.}~\bibnamefont
  {Uppu}}, \bibinfo {author} {\bibfnamefont {F.~T.}\ \bibnamefont {Pedersen}},
  \bibinfo {author} {\bibfnamefont {Y.}~\bibnamefont {Wang}}, \bibinfo {author}
  {\bibfnamefont {C.~T.}\ \bibnamefont {Olesen}}, \bibinfo {author}
  {\bibfnamefont {C.}~\bibnamefont {Papon}}, \bibinfo {author} {\bibfnamefont
  {X.}~\bibnamefont {Zhou}}, \bibinfo {author} {\bibfnamefont {L.}~\bibnamefont
  {Midolo}}, \bibinfo {author} {\bibfnamefont {S.}~\bibnamefont {Scholz}},
  \bibinfo {author} {\bibfnamefont {A.~D.}\ \bibnamefont {Wieck}}, \bibinfo
  {author} {\bibfnamefont {A.}~\bibnamefont {Ludwig}},\ and\ \bibinfo {author}
  {\bibfnamefont {P.}~\bibnamefont {Lodahl}},\ }\href@noop {} {} (\bibinfo
  {year} {2020}),\ \Eprint {http://arxiv.org/abs/2003.08919} {arXiv:2003.08919} \BibitemShut {NoStop}%
\bibitem [{\citenamefont {Wang}\ \emph {et~al.}(2019)\citenamefont {Wang},
  \citenamefont {Qin}, \citenamefont {Ding}, \citenamefont {Chen},
  \citenamefont {Chen}, \citenamefont {You}, \citenamefont {He}, \citenamefont
  {Jiang}, \citenamefont {You}, \citenamefont {Wang}, \citenamefont
  {Schneider}, \citenamefont {Renema}, \citenamefont {H\"ofling}, \citenamefont
  {Lu},\ and\ \citenamefont {Pan}}]{Wang_2019}%
  \BibitemOpen
  \bibfield  {author} {\bibinfo {author} {\bibfnamefont {H.}~\bibnamefont
  {Wang}}, \bibinfo {author} {\bibfnamefont {J.}~\bibnamefont {Qin}}, \bibinfo
  {author} {\bibfnamefont {X.}~\bibnamefont {Ding}}, \bibinfo {author}
  {\bibfnamefont {M.-C.}\ \bibnamefont {Chen}}, \bibinfo {author}
  {\bibfnamefont {S.}~\bibnamefont {Chen}}, \bibinfo {author} {\bibfnamefont
  {X.}~\bibnamefont {You}}, \bibinfo {author} {\bibfnamefont {Y.-M.}\
  \bibnamefont {He}}, \bibinfo {author} {\bibfnamefont {X.}~\bibnamefont
  {Jiang}}, \bibinfo {author} {\bibfnamefont {L.}~\bibnamefont {You}}, \bibinfo
  {author} {\bibfnamefont {Z.}~\bibnamefont {Wang}}, \bibinfo {author}
  {\bibfnamefont {C.}~\bibnamefont {Schneider}}, \bibinfo {author}
  {\bibfnamefont {J.~J.}\ \bibnamefont {Renema}}, \bibinfo {author}
  {\bibfnamefont {S.}~\bibnamefont {H\"ofling}}, \bibinfo {author}
  {\bibfnamefont {C.-Y.}\ \bibnamefont {Lu}},\ and\ \bibinfo {author}
  {\bibfnamefont {J.-W.}\ \bibnamefont {Pan}},\ }\href {\doibase
  10.1103/PhysRevLett.123.250503} {\bibfield  {journal} {\bibinfo  {journal}
  {Phys. Rev. Lett.}\ }\textbf {\bibinfo {volume} {123}},\ \bibinfo {pages}
  {250503} (\bibinfo {year} {2019})}\BibitemShut {NoStop}%
\bibitem [{\citenamefont {Lodahl}(2017)}]{Lodahl_2017}%
  \BibitemOpen
  \bibfield  {author} {\bibinfo {author} {\bibfnamefont {P.}~\bibnamefont
  {Lodahl}},\ }\href {\doibase 10.1088/2058-9565/aa91bb} {\bibfield  {journal}
  {\bibinfo  {journal} {Quantum Sci. Technol.}\ }\textbf {\bibinfo
  {volume} {3}},\ \bibinfo {pages} {013001} (\bibinfo {year}
  {2017})}\BibitemShut {NoStop}%
\bibitem [{\citenamefont {Chang}\ \emph {et~al.}(2018)\citenamefont {Chang},
  \citenamefont {Douglas}, \citenamefont {Gonz\'{a}lez-Tudela}, \citenamefont
  {Hung},\ and\ \citenamefont {Kimble}}]{chang_colloquium:_2018}%
  \BibitemOpen
  \bibfield  {author} {\bibinfo {author} {\bibfnamefont {D.E.}~\bibnamefont
  {Chang}}, \bibinfo {author} {\bibfnamefont {J.S.}~\bibnamefont {Douglas}},
  \bibinfo {author} {\bibfnamefont {A.}~\bibnamefont {Gonz\'{a}lez-Tudela}},
  \bibinfo {author} {\bibfnamefont {C.-L.}\ \bibnamefont {Hung}},\ and\
  \bibinfo {author} {\bibfnamefont {H.J.}~\bibnamefont {Kimble}},\ }\href
  {\doibase 10.1103/RevModPhys.90.031002} {\bibfield  {journal} {\bibinfo
  {journal} {Rev. Mod. Phys.}\ }\textbf {\bibinfo {volume} {90}},\ \bibinfo
  {pages} {031002} (\bibinfo {year} {2018})}\BibitemShut {NoStop}%
\bibitem [{\citenamefont {Arcari}\ \emph {et~al.}(2014)\citenamefont {Arcari},
  \citenamefont {S\"{o}llner}, \citenamefont {Javadi}, \citenamefont
  {Lindskov~Hansen}, \citenamefont {Mahmoodian}, \citenamefont {Liu},
  \citenamefont {Thyrrestrup}, \citenamefont {Lee}, \citenamefont {Song},
  \citenamefont {Stobbe},\ and\ \citenamefont {Lodahl}}]{Arcari_2014}%
  \BibitemOpen
  \bibfield  {author} {\bibinfo {author} {\bibfnamefont {M.}~\bibnamefont
  {Arcari}}, \bibinfo {author} {\bibfnamefont {I.}~\bibnamefont {S\"{o}llner}},
  \bibinfo {author} {\bibfnamefont {A.}~\bibnamefont {Javadi}}, \bibinfo
  {author} {\bibfnamefont {S.}~\bibnamefont {Lindskov~Hansen}}, \bibinfo
  {author} {\bibfnamefont {S.}~\bibnamefont {Mahmoodian}}, \bibinfo {author}
  {\bibfnamefont {J.}~\bibnamefont {Liu}}, \bibinfo {author} {\bibfnamefont
  {H.}~\bibnamefont {Thyrrestrup}}, \bibinfo {author} {\bibfnamefont {E.~H.}\
  \bibnamefont {Lee}}, \bibinfo {author} {\bibfnamefont {J.~D.}\ \bibnamefont
  {Song}}, \bibinfo {author} {\bibfnamefont {S.}~\bibnamefont {Stobbe}},\ and\
  \bibinfo {author} {\bibfnamefont {P.}~\bibnamefont {Lodahl}},\ }\href
  {\doibase 10.1103/PhysRevLett.113.093603} {\bibfield  {journal} {\bibinfo
  {journal} {Phys. Rev. Lett.}\ }\textbf {\bibinfo {volume} {113}},\ \bibinfo
  {pages} {093603} (\bibinfo {year} {2014})}\BibitemShut {NoStop}%
\bibitem [{\citenamefont {Kuhlmann}\ \emph {et~al.}(2015)\citenamefont
  {Kuhlmann}, \citenamefont {Prechtel}, \citenamefont {Houel}, \citenamefont
  {Ludwig}, \citenamefont {Reuter}, \citenamefont {Wieck},\ and\ \citenamefont
  {Warburton}}]{Kuhlmann_2015}%
  \BibitemOpen
  \bibfield  {author} {\bibinfo {author} {\bibfnamefont {A.~V.}\ \bibnamefont
  {Kuhlmann}}, \bibinfo {author} {\bibfnamefont {J.~H.}\ \bibnamefont
  {Prechtel}}, \bibinfo {author} {\bibfnamefont {J.}~\bibnamefont {Houel}},
  \bibinfo {author} {\bibfnamefont {A.}~\bibnamefont {Ludwig}}, \bibinfo
  {author} {\bibfnamefont {D.}~\bibnamefont {Reuter}}, \bibinfo {author}
  {\bibfnamefont {A.~D.}\ \bibnamefont {Wieck}},\ and\ \bibinfo {author}
  {\bibfnamefont {R.~J.}\ \bibnamefont {Warburton}},\ }\href {\doibase
  10.1038/ncomms9204} {\bibfield  {journal} {\bibinfo  {journal} {Nat. Commun.}\ }\textbf {\bibinfo {volume} {6}},\ \bibinfo
  {pages} {8204} (\bibinfo {year} {2015})}\BibitemShut {NoStop}%
\bibitem [{\citenamefont {Pedersen}\ \emph {et~al.}(2020)\citenamefont
  {Pedersen}, \citenamefont {Wang}, \citenamefont {Olesen}, \citenamefont
  {Scholz}, \citenamefont {Wieck}, \citenamefont {Ludwig}, \citenamefont
  {L\"{o}bl}, \citenamefont {Warburton}, \citenamefont {Midolo}, \citenamefont
  {Uppu},\ and\ \citenamefont {Lodahl}}]{Pedersen_2020}%
  \BibitemOpen
  \bibfield  {author} {\bibinfo {author} {\bibfnamefont {F.~T.}\ \bibnamefont
  {Pedersen}}, \bibinfo {author} {\bibfnamefont {Y.}~\bibnamefont {Wang}},
  \bibinfo {author} {\bibfnamefont {C.~T.}\ \bibnamefont {Olesen}}, \bibinfo
  {author} {\bibfnamefont {S.}~\bibnamefont {Scholz}}, \bibinfo {author}
  {\bibfnamefont {A.~D.}\ \bibnamefont {Wieck}}, \bibinfo {author}
  {\bibfnamefont {A.}~\bibnamefont {Ludwig}}, \bibinfo {author} {\bibfnamefont
  {M.~C.}\ \bibnamefont {L\"{o}bl}}, \bibinfo {author} {\bibfnamefont {R.~J.}\
  \bibnamefont {Warburton}}, \bibinfo {author} {\bibfnamefont {L.}~\bibnamefont
  {Midolo}}, \bibinfo {author} {\bibfnamefont {R.}~\bibnamefont {Uppu}},\ and\
  \bibinfo {author} {\bibfnamefont {P.}~\bibnamefont {Lodahl}},\ }
  \href {\doibase 10.1021/acsphotonics.0c00758} {\bibfield  {journal} {\bibinfo  {journal} {ACS Photonics}\ }\textbf {\bibinfo {volume} {7}},\ \bibinfo {number} {9},\ \bibinfo
  {pages} {2343-2349} (\bibinfo {year} {2020})}\BibitemShut {NoStop}%
\bibitem [{\citenamefont {Roy}\ \emph {et~al.}(2017)\citenamefont {Roy},
  \citenamefont {Wilson},\ and\ \citenamefont
  {Firstenberg}}]{roy_colloquium:_2017}%
  \BibitemOpen
  \bibfield  {author} {\bibinfo {author} {\bibfnamefont {D.}~\bibnamefont
  {Roy}}, \bibinfo {author} {\bibfnamefont {C.M.}~\bibnamefont {Wilson}},\ and\
  \bibinfo {author} {\bibfnamefont {O.}~\bibnamefont {Firstenberg}},\ }\href
  {\doibase 10.1103/RevModPhys.89.021001} {\bibfield  {journal} {\bibinfo
  {journal} {Rev. Mod. Phys.}\ }\textbf {\bibinfo {volume} {89}},\ \bibinfo
  {pages} {021001} (\bibinfo {year} {2017})}\BibitemShut {NoStop}%
\bibitem [{\citenamefont {S\'{a}nchez-Burillo}\ \emph
  {et~al.}(2015)\citenamefont {S\'{a}nchez-Burillo}, \citenamefont
  {Garc\'{i}a-Ripoll}, \citenamefont {Mart\'{i}n-Moreno},\ and\ \citenamefont
  {Zueco}}]{sanchez-burillo_nonlinear_2015}%
  \BibitemOpen
  \bibfield  {author} {\bibinfo {author} {\bibfnamefont {E.}~\bibnamefont
  {S\'{a}nchez-Burillo}}, \bibinfo {author} {\bibfnamefont {J.}~\bibnamefont
  {Garc\'{i}a-Ripoll}}, \bibinfo {author} {\bibfnamefont {L.}~\bibnamefont
  {Mart\'{i}n-Moreno}},\ and\ \bibinfo {author} {\bibfnamefont
  {D.}~\bibnamefont {Zueco}},\ }\href {\doibase 10.1039/C4FD00206G} {\bibfield
  {journal} {\bibinfo  {journal} {Faraday Discuss.}\ }\textbf {\bibinfo
  {volume} {178}},\ \bibinfo {pages} {335} (\bibinfo {year}
  {2015})}\BibitemShut {NoStop}%
\bibitem [{\citenamefont {Shen}\ and\ \citenamefont
  {Fan}(2007{\natexlab{a}})}]{Shen_2007multi}%
  \BibitemOpen
  \bibfield  {author} {\bibinfo {author} {\bibfnamefont {J.-T.}\ \bibnamefont
  {Shen}}\ and\ \bibinfo {author} {\bibfnamefont {S.}~\bibnamefont {Fan}},\
  }\href {\doibase 10.1103/PhysRevA.76.062709} {\bibfield  {journal} {\bibinfo
  {journal} {Phys. Rev. A}\ }\textbf {\bibinfo {volume} {76}},\ \bibinfo
  {pages} {062709} (\bibinfo {year} {2007}{\natexlab{a}})}\BibitemShut
  {NoStop}%
\bibitem [{\citenamefont {Shen}\ and\ \citenamefont
  {Fan}(2007{\natexlab{b}})}]{Shen_2010}%
  \BibitemOpen
  \bibfield  {author} {\bibinfo {author} {\bibfnamefont {J.-T.}\ \bibnamefont
  {Shen}}\ and\ \bibinfo {author} {\bibfnamefont {S.}~\bibnamefont {Fan}},\
  }\href {\doibase 10.1103/PhysRevLett.98.153003} {\bibfield  {journal}
  {\bibinfo  {journal} {Phys. Rev. Lett.}\ }\textbf {\bibinfo {volume} {98}},\
  \bibinfo {pages} {153003} (\bibinfo {year} {2007}{\natexlab{b}})}\BibitemShut
  {NoStop}%
\bibitem [{\citenamefont {Fan}\ \emph {et~al.}(2010)\citenamefont {Fan},
  \citenamefont {Kocaba\c{s}},\ and\ \citenamefont {Shen}}]{Fan_2010}%
  \BibitemOpen
  \bibfield  {author} {\bibinfo {author} {\bibfnamefont {S.}~\bibnamefont
  {Fan}}, \bibinfo {author} {\bibfnamefont {S.~E.}\ \bibnamefont
  {Kocaba\c{s}}},\ and\ \bibinfo {author} {\bibfnamefont {J.-T.}\ \bibnamefont
  {Shen}},\ }\href {\doibase 10.1103/PhysRevA.82.063821} {\bibfield  {journal}
  {\bibinfo  {journal} {Phys. Rev. A}\ }\textbf {\bibinfo {volume} {82}},\
  \bibinfo {pages} {063821} (\bibinfo {year} {2010})}\BibitemShut {NoStop}%
\bibitem [{\citenamefont {Das}\ \emph {et~al.}(2018)\citenamefont {Das},
  \citenamefont {Elfving}, \citenamefont {Reiter},\ and\ \citenamefont
  {S\o{}rensen}}]{das_photon_2018}%
  \BibitemOpen
  \bibfield  {author} {\bibinfo {author} {\bibfnamefont {S.}~\bibnamefont
  {Das}}, \bibinfo {author} {\bibfnamefont {V.~E.}\ \bibnamefont {Elfving}},
  \bibinfo {author} {\bibfnamefont {F.}~\bibnamefont {Reiter}},\ and\ \bibinfo
  {author} {\bibfnamefont {A.~S.}\ \bibnamefont {S\o{}rensen}},\ }\href
  {\doibase 10.1103/PhysRevA.97.043838} {\bibfield  {journal} {\bibinfo
  {journal} {Phys. Rev. A}\ }\textbf {\bibinfo {volume} {97}},\ \bibinfo
  {pages} {043838} (\bibinfo {year} {2018})}\BibitemShut {NoStop}%
\bibitem [{\citenamefont {Ramos}\ and\ \citenamefont
  {Garc{\'{\i}}a-Ripoll}(2018)}]{Ramos_2018}%
  \BibitemOpen
  \bibfield  {author} {\bibinfo {author} {\bibfnamefont {T.}~\bibnamefont
  {Ramos}}\ and\ \bibinfo {author} {\bibfnamefont {J.~J.}\ \bibnamefont
  {Garc{\'{\i}}a-Ripoll}},\ }\href {\doibase 10.1088/1367-2630/aae73b}
  {\bibfield  {journal} {\bibinfo  {journal} {New J. Phys.}\ }\textbf
  {\bibinfo {volume} {20}},\ \bibinfo {pages} {105007} (\bibinfo {year}
  {2018})}\BibitemShut {NoStop}%
\bibitem [{\citenamefont {Trivedi}\ \emph {et~al.}(2018)\citenamefont
  {Trivedi}, \citenamefont {Fischer}, \citenamefont {Xu}, \citenamefont {Fan},\
  and\ \citenamefont {Vu{\v c}kovi{\'c}}}]{trivedi_few-photon_2018}%
  \BibitemOpen
  \bibfield  {author} {\bibinfo {author} {\bibfnamefont {R.}~\bibnamefont
  {Trivedi}}, \bibinfo {author} {\bibfnamefont {K.}~\bibnamefont {Fischer}},
  \bibinfo {author} {\bibfnamefont {S.}~\bibnamefont {Xu}}, \bibinfo {author}
  {\bibfnamefont {S.}~\bibnamefont {Fan}},\ and\ \bibinfo {author}
  {\bibfnamefont {J.}~\bibnamefont {Vu{\v c}kovi{\'c}}},\ }\href {\doibase
  10.1103/PhysRevB.98.144112} {\bibfield  {journal} {\bibinfo  {journal} {Phys.
  Rev. B}\ }\textbf {\bibinfo {volume} {98}},\ \bibinfo {pages} {144112}
  (\bibinfo {year} {2018})}\BibitemShut {NoStop}%
\bibitem [{\citenamefont {Zheng}\ \emph {et~al.}(2010)\citenamefont {Zheng},
  \citenamefont {Gauthier},\ and\ \citenamefont
  {Baranger}}]{zheng_waveguide_2010}%
  \BibitemOpen
  \bibfield  {author} {\bibinfo {author} {\bibfnamefont {H.}~\bibnamefont
  {Zheng}}, \bibinfo {author} {\bibfnamefont {D.~J.}\ \bibnamefont {Gauthier}},
 \ and\ \bibinfo {author} {\bibfnamefont {H.~U.}\ \bibnamefont {Baranger}},\
  }\href {\doibase 10.1103/PhysRevA.82.063816} {\bibfield  {journal} {\bibinfo
  {journal} {Phys. Rev. A}\ }\textbf {\bibinfo {volume} {82}},\ \bibinfo
  {pages} {063816} (\bibinfo {year} {2010})}\BibitemShut {NoStop}%
\bibitem [{\citenamefont {Huang}\ \emph {et~al.}(2013)\citenamefont {Huang},
  \citenamefont {Shi}, \citenamefont {Sun},\ and\ \citenamefont
  {Nori}}]{huang_controlling_2013}%
  \BibitemOpen
  \bibfield  {author} {\bibinfo {author} {\bibfnamefont {J.-F.}\ \bibnamefont
  {Huang}}, \bibinfo {author} {\bibfnamefont {T.}~\bibnamefont {Shi}}, \bibinfo
  {author} {\bibfnamefont {C.~P.}\ \bibnamefont {Sun}},\ and\ \bibinfo
  {author} {\bibfnamefont {F.}~\bibnamefont {Nori}},\ }\href {\doibase
  10.1103/PhysRevA.88.013836} {\bibfield  {journal} {\bibinfo  {journal} {Phys.
  Rev. A}\ }\textbf {\bibinfo {volume} {88}},\ \bibinfo {pages} {013836}
  (\bibinfo {year} {2013})},\ \bibinfo {note} {publisher: American Physical
  Society}\BibitemShut {NoStop}%
\bibitem [{\citenamefont {Lee}\ \emph {et~al.}(2015)\citenamefont {Lee},
  \citenamefont {Noh}, \citenamefont {Schetakis},\ and\ \citenamefont
  {Angelakis}}]{lee_few-photon_2015}%
  \BibitemOpen
  \bibfield  {author} {\bibinfo {author} {\bibfnamefont {C.}~\bibnamefont
  {Lee}}, \bibinfo {author} {\bibfnamefont {C.}~\bibnamefont {Noh}}, \bibinfo
  {author} {\bibfnamefont {N.}~\bibnamefont {Schetakis}},\ and\ \bibinfo
  {author} {\bibfnamefont {D.~G.}\ \bibnamefont {Angelakis}},\ }\href {\doibase
  10.1103/PhysRevA.92.063817} {\bibfield  {journal} {\bibinfo  {journal} {Phys.
  Rev. A}\ }\textbf {\bibinfo {volume} {92}},\ \bibinfo {pages} {063817}
  (\bibinfo {year} {2015})}\BibitemShut {NoStop}%
\bibitem [{\citenamefont {Das}\ \emph {et~al.}(2019)\citenamefont {Das},
  \citenamefont {Zhai}, \citenamefont {{\v C}epulskovskis}, \citenamefont
  {Javadi}, \citenamefont {Mahmoodian}, \citenamefont {Lodahl},\ and\
  \citenamefont {S\o{}rensen}}]{das_wave-function_2019}%
  \BibitemOpen
  \bibfield  {author} {\bibinfo {author} {\bibfnamefont {S.}~\bibnamefont
  {Das}}, \bibinfo {author} {\bibfnamefont {L.}~\bibnamefont {Zhai}}, \bibinfo
  {author} {\bibfnamefont {M.}~\bibnamefont {{\v C}epulskovskis}}, \bibinfo
  {author} {\bibfnamefont {A.}~\bibnamefont {Javadi}}, \bibinfo {author}
  {\bibfnamefont {S.}~\bibnamefont {Mahmoodian}}, \bibinfo {author}
  {\bibfnamefont {P.}~\bibnamefont {Lodahl}},\ and\ \bibinfo {author}
  {\bibfnamefont {A.~S.}\ \bibnamefont {S\o{}rensen}},\ }\href
  {http://arxiv.org/abs/1912.08303} {\bibfield  {journal} {\bibinfo  {journal}
  {arXiv:1912.08303}\ } (\bibinfo {year}
  {2019})}\BibitemShut {NoStop}%
\bibitem [{\citenamefont {Pedersen}\ and\ \citenamefont
  {Pletyukhov}(2017)}]{pedersen_few-photon_2017}%
  \BibitemOpen
  \bibfield  {author} {\bibinfo {author} {\bibfnamefont {K.~G.~L.}\
  \bibnamefont {Pedersen}}\ and\ \bibinfo {author} {\bibfnamefont
  {M.}~\bibnamefont {Pletyukhov}},\ }\href {\doibase
  10.1103/PhysRevA.96.023815} {\bibfield  {journal} {\bibinfo  {journal} {Phys.
  Rev. A}\ }\textbf {\bibinfo {volume} {96}},\ \bibinfo {pages} {023815}
  (\bibinfo {year} {2017})}\BibitemShut {NoStop}%
\bibitem [{\citenamefont {Lang}\ \emph {et~al.}(2018)\citenamefont {Lang},
  \citenamefont {Chang},\ and\ \citenamefont
  {Piazza}}]{lang_non-equilibrium_2018}%
  \BibitemOpen
  \bibfield  {author} {\bibinfo {author} {\bibfnamefont {J.}~\bibnamefont
  {Lang}}, \bibinfo {author} {\bibfnamefont {D.~E.}\ \bibnamefont {Chang}}, \
  and\ \bibinfo {author} {\bibfnamefont {F.}~\bibnamefont {Piazza}},\ }\href
  {http://arxiv.org/abs/1810.12921} {\bibfield  {journal} {\bibinfo  {journal}
  {arXiv:1810.12921}\ } (\bibinfo {year}
  {2018})}\BibitemShut {NoStop}%
\bibitem [{\citenamefont {Shi}\ \emph {et~al.}(2015)\citenamefont {Shi},
  \citenamefont {Chang},\ and\ \citenamefont
  {Cirac}}]{shi_multiphoton-scattering_2015}%
  \BibitemOpen
  \bibfield  {author} {\bibinfo {author} {\bibfnamefont {T.}~\bibnamefont
  {Shi}}, \bibinfo {author} {\bibfnamefont {D.~E.}\ \bibnamefont {Chang}}, \
  and\ \bibinfo {author} {\bibfnamefont {J.~I.}\ \bibnamefont {Cirac}},\ }\href
  {\doibase 10.1103/PhysRevA.92.053834} {\bibfield  {journal} {\bibinfo
  {journal} {Phys. Rev. A}\ }\textbf {\bibinfo {volume} {92}},\ \bibinfo
  {pages} {053834} (\bibinfo {year} {2015})}\BibitemShut {NoStop}%
\bibitem [{\citenamefont {Shi}\ \emph {et~al.}(2018)\citenamefont {Shi},
  \citenamefont {Chang},\ and\ \citenamefont
  {Garc\'{i}a-Ripoll}}]{shi_ultrastrong_2018}%
  \BibitemOpen
  \bibfield  {author} {\bibinfo {author} {\bibfnamefont {T.}~\bibnamefont
  {Shi}}, \bibinfo {author} {\bibfnamefont {Y.}~\bibnamefont {Chang}},\ and\
  \bibinfo {author} {\bibfnamefont {J.~J.}\ \bibnamefont {Garc\'{i}a-Ripoll}},\
  }\href {\doibase 10.1103/PhysRevLett.120.153602} {\bibfield  {journal}
  {\bibinfo  {journal} {Phys. Rev. Lett.}\ }\textbf {\bibinfo {volume} {120}},\
  \bibinfo {pages} {153602} (\bibinfo {year} {2018})}\BibitemShut {NoStop}%
\bibitem [{\citenamefont {Bera}\ \emph {et~al.}(2014)\citenamefont {Bera},
  \citenamefont {Nazir}, \citenamefont {Chin}, \citenamefont {Baranger},\ and\
  \citenamefont {Florens}}]{bera_generalized_2014}%
  \BibitemOpen
  \bibfield  {author} {\bibinfo {author} {\bibfnamefont {S.}~\bibnamefont
  {Bera}}, \bibinfo {author} {\bibfnamefont {A.}~\bibnamefont {Nazir}},
  \bibinfo {author} {\bibfnamefont {A.~W.}\ \bibnamefont {Chin}}, \bibinfo
  {author} {\bibfnamefont {H.~U.}\ \bibnamefont {Baranger}},\ and\ \bibinfo
  {author} {\bibfnamefont {S.}~\bibnamefont {Florens}},\ }\href {\doibase
  10.1103/PhysRevB.90.075110} {\bibfield  {journal} {\bibinfo  {journal} {Phys.
  Rev. B}\ }\textbf {\bibinfo {volume} {90}},\ \bibinfo {pages} {075110}
  (\bibinfo {year} {2014})},\BibitemShut {NoStop}%
\bibitem [{\citenamefont {Witthaut}\ \emph {et~al.}(2012)\citenamefont
  {Witthaut}, \citenamefont {Lukin},\ and\ \citenamefont
  {S\o{}rensen}}]{Witthaut_2012}%
  \BibitemOpen
  \bibfield  {author} {\bibinfo {author} {\bibfnamefont {D.}~\bibnamefont
  {Witthaut}}, \bibinfo {author} {\bibfnamefont {M.~D.}\ \bibnamefont {Lukin}},
 \ and\ \bibinfo {author} {\bibfnamefont {A.~S.}\ \bibnamefont
  {S\o{}rensen}},\ }\href {\doibase 10.1209/0295-5075/97/50007} {\bibfield
  {journal} {\bibinfo  {journal} {EPL (Europhysics Letters)}\ }\textbf
  {\bibinfo {volume} {97}},\ \bibinfo {pages} {50007} (\bibinfo {year}
  {2012})}\BibitemShut {NoStop}%
\bibitem [{\citenamefont {Ralph}\ \emph {et~al.}(2015)\citenamefont {Ralph},
  \citenamefont {S\"ollner}, \citenamefont {Mahmoodian}, \citenamefont
  {White},\ and\ \citenamefont {Lodahl}}]{Ralph_2015}%
  \BibitemOpen
  \bibfield  {author} {\bibinfo {author} {\bibfnamefont {T.~C.}\ \bibnamefont
  {Ralph}}, \bibinfo {author} {\bibfnamefont {I.}~\bibnamefont {S\"ollner}},
  \bibinfo {author} {\bibfnamefont {S.}~\bibnamefont {Mahmoodian}}, \bibinfo
  {author} {\bibfnamefont {A.~G.}\ \bibnamefont {White}},\ and\ \bibinfo
  {author} {\bibfnamefont {P.}~\bibnamefont {Lodahl}},\ }\href {\doibase
  10.1103/PhysRevLett.114.173603} {\bibfield  {journal} {\bibinfo  {journal}
  {Phys. Rev. Lett.}\ }\textbf {\bibinfo {volume} {114}},\ \bibinfo {pages}
  {173603} (\bibinfo {year} {2015})}\BibitemShut {NoStop}%
\bibitem [{\citenamefont {Firstenberg}\ \emph {et~al.}(2013)\citenamefont
  {Firstenberg}, \citenamefont {Peyronel}, \citenamefont {Liang}, \citenamefont
  {Gorshkov}, \citenamefont {Lukin},\ and\ \citenamefont
  {Vuleti{\'{c}}}}]{firstenberg_attractive_2013}%
  \BibitemOpen
  \bibfield  {author} {\bibinfo {author} {\bibfnamefont {O.}~\bibnamefont
  {Firstenberg}}, \bibinfo {author} {\bibfnamefont {T.}~\bibnamefont
  {Peyronel}}, \bibinfo {author} {\bibfnamefont {Q.-Y.}\ \bibnamefont {Liang}},
  \bibinfo {author} {\bibfnamefont {A.~V.}\ \bibnamefont {Gorshkov}}, \bibinfo
  {author} {\bibfnamefont {M.~D.}\ \bibnamefont {Lukin}},\ and\ \bibinfo
  {author} {\bibfnamefont {V.}~\bibnamefont {Vuleti{\'{c}}}},\ }\href {\doibase
  10.1038/nature12512} {\bibfield  {journal} {\bibinfo  {journal} {Nature}\
  }\textbf {\bibinfo {volume} {502}},\ \bibinfo {pages} {71} (\bibinfo {year}
  {2013})}\BibitemShut {NoStop}%
\bibitem [{\citenamefont {Mahmoodian}\ \emph {et~al.}(2019)\citenamefont
  {Mahmoodian}, \citenamefont {Calaj\'{o}}, \citenamefont {Chang},
  \citenamefont {Hammerer},\ and\ \citenamefont
  {S\o{}rensen}}]{Mahmoodian_2019}%
  \BibitemOpen
  \bibfield  {author} {\bibinfo {author} {\bibfnamefont {S.}~\bibnamefont
  {Mahmoodian}}, \bibinfo {author} {\bibfnamefont {G.}~\bibnamefont
  {Calaj\'{o}}}, \bibinfo {author} {\bibfnamefont {D.~E.}\ \bibnamefont
  {Chang}}, \bibinfo {author} {\bibfnamefont {K.}~\bibnamefont {Hammerer}}, \
  and\ \bibinfo {author} {\bibfnamefont {A.~S.}\ \bibnamefont {S\o{}rensen}},\
  }\href@noop {} {} (\bibinfo {year} {2019}),\ \Eprint
  {http://arxiv.org/abs/1910.05828} {arXiv:1910.05828} \BibitemShut {NoStop}%
\bibitem [{\citenamefont {Ramos}\ and\ \citenamefont
  {Garc\'{\i}a-Ripoll}(2017)}]{Ramos_2017}%
  \BibitemOpen
  \bibfield  {author} {\bibinfo {author} {\bibfnamefont {T.}~\bibnamefont
  {Ramos}}\ and\ \bibinfo {author} {\bibfnamefont {J.~J.}\ \bibnamefont
  {Garc\'{\i}a-Ripoll}},\ }\href {\doibase 10.1103/PhysRevLett.119.153601}
  {\bibfield  {journal} {\bibinfo  {journal} {Phys. Rev. Lett.}\ }\textbf
  {\bibinfo {volume} {119}},\ \bibinfo {pages} {153601} (\bibinfo {year}
  {2017})}\BibitemShut {NoStop}%
\bibitem [{\citenamefont {T\"{u}rschmann}\ \emph {et~al.}(2019)\citenamefont
  {T\"{u}rschmann}, \citenamefont {Jeannic}, \citenamefont {Simonsen},
  \citenamefont {Haakh}, \citenamefont {G\"{o}tzinger}, \citenamefont
  {Sandoghdar}, \citenamefont {Lodahl},\ and\ \citenamefont
  {Rotenberg}}]{Turschmann_2019}%
  \BibitemOpen
  \bibfield  {author} {\bibinfo {author} {\bibfnamefont {P.}~\bibnamefont
  {T\"{u}rschmann}}, \bibinfo {author} {\bibfnamefont {H.~L.}\ \bibnamefont
  {Jeannic}}, \bibinfo {author} {\bibfnamefont {S.~F.}\ \bibnamefont
  {Simonsen}}, \bibinfo {author} {\bibfnamefont {H.~R.}\ \bibnamefont {Haakh}},
  \bibinfo {author} {\bibfnamefont {S.}~\bibnamefont {G\"{o}tzinger}}, \bibinfo
  {author} {\bibfnamefont {V.}~\bibnamefont {Sandoghdar}}, \bibinfo {author}
  {\bibfnamefont {P.}~\bibnamefont {Lodahl}},\ and\ \bibinfo {author}
  {\bibfnamefont {N.}~\bibnamefont {Rotenberg}},\ }\href {\doibase
  https://doi.org/10.1515/nanoph-2019-0126} {\bibfield  {journal} {\bibinfo
  {journal} {Nanophotonics}\ }\textbf {\bibinfo {volume} {8}},\ \bibinfo
  {pages} {1641 } (\bibinfo {year} {2019})}\BibitemShut {NoStop}%
\bibitem [{\citenamefont {Lodahl}\ \emph {et~al.}(2017)\citenamefont {Lodahl},
  \citenamefont {Mahmoodian}, \citenamefont {Stobbe}, \citenamefont
  {Rauschenbeutel}, \citenamefont {Schneeweiss}, \citenamefont {Volz},
  \citenamefont {Pichler},\ and\ \citenamefont {Zoller}}]{Lodahl2017_Chiral}%
  \BibitemOpen
  \bibfield  {author} {\bibinfo {author} {\bibfnamefont {P.}~\bibnamefont
  {Lodahl}}, \bibinfo {author} {\bibfnamefont {S.}~\bibnamefont {Mahmoodian}},
  \bibinfo {author} {\bibfnamefont {S.}~\bibnamefont {Stobbe}}, \bibinfo
  {author} {\bibfnamefont {A.}~\bibnamefont {Rauschenbeutel}}, \bibinfo
  {author} {\bibfnamefont {P.}~\bibnamefont {Schneeweiss}}, \bibinfo {author}
  {\bibfnamefont {J.}~\bibnamefont {Volz}}, \bibinfo {author} {\bibfnamefont
  {H.}~\bibnamefont {Pichler}},\ and\ \bibinfo {author} {\bibfnamefont
  {P.}~\bibnamefont {Zoller}},\ }\href {\doibase
  https://doi.org/10.1038/nature21037} {\bibfield  {journal} {\bibinfo
  {journal} {Nature}\ }\textbf {\bibinfo {volume} {541}},\ \bibinfo {pages}
  {473–480} (\bibinfo {year} {2017})}\BibitemShut {NoStop}%
\bibitem [{\citenamefont {Ramos}\ \emph {et~al.}(2014)\citenamefont {Ramos},
  \citenamefont {Pichler}, \citenamefont {Daley},\ and\ \citenamefont
  {Zoller}}]{ramos_quantum_2014}%
  \BibitemOpen
  \bibfield  {author} {\bibinfo {author} {\bibfnamefont {T.}~\bibnamefont
  {Ramos}}, \bibinfo {author} {\bibfnamefont {H.}~\bibnamefont {Pichler}},
  \bibinfo {author} {\bibfnamefont {A.~J.}\ \bibnamefont {Daley}},\ and\
  \bibinfo {author} {\bibfnamefont {P.}~\bibnamefont {Zoller}},\ }\href
  {\doibase 10.1103/PhysRevLett.113.237203} {\bibfield  {journal} {\bibinfo
  {journal} {Phys. Rev. Lett.}\ }\textbf {\bibinfo {volume} {113}},\ \bibinfo
  {pages} {237203} (\bibinfo {year} {2014})}\BibitemShut {NoStop}%
\bibitem [{SM()}]{SM}%
  \BibitemOpen
  \href@noop {} {}\bibinfo {note} {See Supplemental Material [URL] for (i) the
  modeling of the QD-waveguide system and measurements including all
  experimental imperfections, and for (ii) the derivation of the two-photon
  reconstruction formulas, which additionnally includes Refs. \cite{Gardiner-Zoller_2004,Auffeves_2007,Blow_1990}.}\BibitemShut {Stop}%
\bibitem [{\citenamefont {Ramos}(tion)}]{Ramos_2020}%
  \BibitemOpen
  \bibfield  {author} {\bibinfo {author} {\bibfnamefont {T.}~\bibnamefont
  {Ramos}},\ }\href@noop {} {} (\bibinfo {year} {In preparation})\BibitemShut
  {NoStop}%
\BibitemOpen
\bibitem [{\citenamefont {Kir{\v{s}}ansk{\.{e}}}\ \emph
  {et~al.}(2017)\citenamefont {Kir{\v{s}}ansk{\.{e}}}, \citenamefont
  {Thyrrestrup}, \citenamefont {Daveau}, \citenamefont {Dree\ss{}en},
  \citenamefont {Pregnolato}, \citenamefont {Midolo}, \citenamefont
  {Tighineanu}, \citenamefont {Javadi}, \citenamefont {Stobbe}, \citenamefont
  {Schott}, \citenamefont {Ludwig}, \citenamefont {Wieck}, \citenamefont
  {Park}, \citenamefont {Song}, \citenamefont {Kuhlmann}, \citenamefont
  {S\"ollner}, \citenamefont {L\"obl}, \citenamefont {Warburton},\ and\
  \citenamefont {Lodahl}}]{Kirsanske2017}%
  \BibitemOpen
  \bibfield  {author} {\bibinfo {author} {\bibfnamefont {G.}~\bibnamefont
  {Kir{\v{s}}ansk{\.{e}}}}, \bibinfo {author} {\bibfnamefont {H.}~\bibnamefont
  {Thyrrestrup}}, \bibinfo {author} {\bibfnamefont {R.~S.}\ \bibnamefont
  {Daveau}}, \bibinfo {author} {\bibfnamefont {C.~L.}\ \bibnamefont
  {Dree\ss{}en}}, \bibinfo {author} {\bibfnamefont {T.}~\bibnamefont
  {Pregnolato}}, \bibinfo {author} {\bibfnamefont {L.}~\bibnamefont {Midolo}},
  \bibinfo {author} {\bibfnamefont {P.}~\bibnamefont {Tighineanu}}, \bibinfo
  {author} {\bibfnamefont {A.}~\bibnamefont {Javadi}}, \bibinfo {author}
  {\bibfnamefont {S.}~\bibnamefont {Stobbe}}, \bibinfo {author} {\bibfnamefont
  {R.}~\bibnamefont {Schott}}, \bibinfo {author} {\bibfnamefont
  {A.}~\bibnamefont {Ludwig}}, \bibinfo {author} {\bibfnamefont {A.~D.}\
  \bibnamefont {Wieck}}, \bibinfo {author} {\bibfnamefont {S.~I.}\ \bibnamefont
  {Park}}, \bibinfo {author} {\bibfnamefont {J.~D.}\ \bibnamefont {Song}},
  \bibinfo {author} {\bibfnamefont {A.~V.}\ \bibnamefont {Kuhlmann}}, \bibinfo
  {author} {\bibfnamefont {I.}~\bibnamefont {S\"ollner}}, \bibinfo {author}
  {\bibfnamefont {M.~C.}\ \bibnamefont {L\"obl}}, \bibinfo {author}
  {\bibfnamefont {R.~J.}\ \bibnamefont {Warburton}},\ and\ \bibinfo {author}
  {\bibfnamefont {P.}~\bibnamefont {Lodahl}},\ }\href {\doibase
  10.1103/PhysRevB.96.165306} {\bibfield  {journal} {\bibinfo  {journal} {Phys.
  Rev. B}\ }\textbf {\bibinfo {volume} {96}},\ \bibinfo {pages} {165306}
  (\bibinfo {year} {2017})}\BibitemShut{NoStop}%
\bibitem [{\citenamefont {Zhou}\ \emph {et~al.}(2018)\citenamefont {Zhou},
  \citenamefont {Kulkova}, \citenamefont {Lund-Hansen}, \citenamefont
  {Lindskov~Hansen}, \citenamefont {Lodahl},\ and\ \citenamefont
  {Midolo}}]{Zhou_2018}
  \BibitemOpen
  \bibfield  {author} {\bibinfo {author} {\bibfnamefont {X.}~\bibnamefont
  {Zhou}}, \bibinfo {author} {\bibfnamefont {I.}~\bibnamefont {Kulkova}},
  \bibinfo {author} {\bibfnamefont {T.}~\bibnamefont {Lund-Hansen}}, \bibinfo
  {author} {\bibfnamefont {S.}~\bibnamefont {Lindskov~Hansen}}, \bibinfo
  {author} {\bibfnamefont {P.}~\bibnamefont {Lodahl}},\ and\ \bibinfo {author}
  {\bibfnamefont {L.}~\bibnamefont {Midolo}},\ }\href {\doibase
  10.1063/1.5055622} {\bibfield  {journal} {\bibinfo  {journal} {App.
  Phys. Lett.}\ }\textbf {\bibinfo {volume} {113}},\ \bibinfo {pages}
  {251103} (\bibinfo {year} {2018})}\BibitemShut{NoStop}%
\bibitem [{\citenamefont {Meissner}(tion)}]{Meissner_2012}%
  \BibitemOpen
  \bibfield  {author} {\bibinfo {author} {\bibfnamefont {M.}~\bibnamefont
  {Meissner}},\ }\href {\doibase
  10.12693/APhysPolA.121.A-164} {\bibfield  {journal} {\bibinfo  {journal} {Acta Physica Polonica A}\ }\textbf {\bibinfo {volume} {121}}, \bibinfo {pages}
  {A-164} (\bibinfo {year} {2012})}\BibitemShut {NoStop}%
\bibitem [{\citenamefont {Hansom}\ \emph {et~al.}(2014)\citenamefont {Hansom},
  \citenamefont {Schulte}, \citenamefont {Matthiesen}, \citenamefont
  {Stanley},\ and\ \citenamefont {Atat\"{u}re}}]{Hansom_2014}%
  \BibitemOpen
  \bibfield  {author} {\bibinfo {author} {\bibfnamefont {J.}~\bibnamefont
  {Hansom}}, \bibinfo {author} {\bibfnamefont {C.~H.~H.}\ \bibnamefont
  {Schulte}}, \bibinfo {author} {\bibfnamefont {C.}~\bibnamefont {Matthiesen}},
  \bibinfo {author} {\bibfnamefont {M.~J.}\ \bibnamefont {Stanley}},\ and\
  \bibinfo {author} {\bibfnamefont {M.}~\bibnamefont {Atat\"{u}re}},\ }\href
  {\doibase 10.1063/1.4901045} {\bibfield  {journal} {\bibinfo  {journal}
  {App. Phys. Lett.}\ }\textbf {\bibinfo {volume} {105}},\ \bibinfo
  {pages} {172107} (\bibinfo {year} {2014})}\BibitemShut {NoStop}%
\bibitem [{\citenamefont {Gardiner}\ and\ \citenamefont
  {Zoller}(2004)}]{Gardiner-Zoller_2004}%
  \BibitemOpen
  \bibfield  {author} {\bibinfo {author} {\bibfnamefont {C.~W.}\ \bibnamefont
  {Gardiner}}\ and\ \bibinfo {author} {\bibfnamefont {P.}~\bibnamefont
  {Zoller}},\ }\href@noop {} {\emph {\bibinfo {title} {Quantum Noise}}}\
  (\bibinfo  {publisher} {Springer-Verlag Berlin Heidelberg},\ \bibinfo {year}
  {2004})\BibitemShut {NoStop}%
\bibitem [{\citenamefont {Auff\`eves-Garnier}\ \emph
  {et~al.}(2007)\citenamefont {Auff\`eves-Garnier}, \citenamefont {Simon},
  \citenamefont {G\'erard},\ and\ \citenamefont {Poizat}}]{Auffeves_2007}%
  \BibitemOpen
  \bibfield  {author} {\bibinfo {author} {\bibfnamefont {A.}~\bibnamefont
  {Auff\`eves-Garnier}}, \bibinfo {author} {\bibfnamefont {C.}~\bibnamefont
  {Simon}}, \bibinfo {author} {\bibfnamefont {J.-M.}\ \bibnamefont
  {G\'erard}},\ and\ \bibinfo {author} {\bibfnamefont {J.-P.}\ \bibnamefont
  {Poizat}},\ }\href {https://doi.org/10.1103/PhysRevA.75.053823} {\bibfield
  {journal} {\bibinfo  {journal} {Phys. Rev. A}\ }\textbf {\bibinfo {volume}
  {75}},\ \bibinfo {pages} {053823} (\bibinfo {year} {2007})}\BibitemShut
  {NoStop}%
\bibitem [{\citenamefont {Blow}\ \emph {et~al.}(1990)\citenamefont {Blow},
  \citenamefont {Loudon}, \citenamefont {Phoenix},\ and\ \citenamefont
  {Shepherd}}]{Blow_1990}%
  \BibitemOpen
  \bibfield  {author} {\bibinfo {author} {\bibfnamefont {K.~J.}\ \bibnamefont
  {Blow}}, \bibinfo {author} {\bibfnamefont {R.}~\bibnamefont {Loudon}},
  \bibinfo {author} {\bibfnamefont {S.~J.~D.}\ \bibnamefont {Phoenix}},\ and\
  \bibinfo {author} {\bibfnamefont {T.~J.}\ \bibnamefont {Shepherd}},\
  }\href {https://doi.org/10.1103/PhysRevA.42.4102} {\bibfield  {journal}
  {\bibinfo  {journal} {Phys. Rev. A}\ }\textbf {\bibinfo {volume} {42}},\
  \bibinfo {pages} {4102} (\bibinfo {year} {1990})}\BibitemShut {NoStop}%
  
  
  
\end{thebibliography}
\end{document}